\documentclass[prl,aps,twocolumn,superscriptaddress,notitlepage,10pt]{revtex4-2}
\usepackage{amssymb,amsfonts,amsmath,amstext,graphicx}
\usepackage{amsmath}
\usepackage{tikz}
\usepackage{graphicx}
\usepackage[normalem]{ulem}
\usepackage{tabularx}
\def\bra#1{\langle{#1}|}
\def\ket#1{|{#1}\rangle}
\def\braket#1{\langle{#1}\rangle}

\def\BraVert{\egroup\,\mid\,\bgroup}

\usepackage{braket}
\usepackage{xcolor}
\usepackage[colorlinks, linkcolor=red,urlcolor=blue,citecolor=blue]{hyperref}

\begin{document}

\title{Universal relaxation speedup in nonequilibrium quantum many-body systems through transient resetting}

\author{Parvinder Solanki} \email{parvinder@mnf.uni-tuebingen.de}
\affiliation{Institut f\"ur Theoretische Physik and Center for Integrated Quantum Science and Technology, Universit\"at T\"ubingen, Auf der Morgenstelle 14, 72076 T\"ubingen, Germany}
\affiliation{Department of Physics, Indian Institute of Space Science and Technology, Thiruvananthapuram, Kerala 695547, India}
\author{Igor Lesanovsky} \affiliation{Institut f\"ur Theoretische Physik and Center for Integrated Quantum Science and Technology, Universit\"at T\"ubingen, Auf der Morgenstelle 14, 72076 T\"ubingen, Germany}
\affiliation{School of Physics and Astronomy and Centre for the Mathematics and Theoretical Physics of Quantum Non-Equilibrium Systems, University of Nottingham, Nottingham, NG7 2RD, United Kingdom}
\author{Gabriele Perfetto}  
\email{gperfetto@phys.ethz.ch}
\affiliation{Institut für Theoretische Physik, ETH Zürich, Wolfgang-Pauli-Str. 27, 8093 Zürich, Switzerland}

\begin{abstract}
Speeding up the relaxation dynamics of many-body quantum systems is important in a variety of contexts, including quantum computation and state preparation. We demonstrate that such acceleration can be universally achieved via transient stochastic resetting. 
This means that during an initial time interval of finite duration, the dynamics is interrupted by resets that take the system to a designated state at randomly selected times. We illustrate this idea for few-body open systems and also for a challenging many-body case, where a first-order phase transition leads to a divergence of relaxation time. 
In all scenarios, a significant and sometimes even exponential acceleration in reaching the stationary state is observed, similar to the so-called Mpemba effect. 
The universal nature of this speedup lies in the fact that the design of the resetting protocol only requires knowledge of a few macroscopic properties of the target state, such as the order parameter of the phase transition, while it does not necessitate any fine-tuned manipulation of the initial state.

\end{abstract}

\maketitle

\paragraph{Introduction.---}
Quantum systems interacting with their environment generically relax to stationary states. For many-body systems, the associated relaxation timescales can become extremely slow, e.g., near first-order phase transitions, leading to metastable behaviors that render stationary state properties practically unreachable \cite{minganti2018dissipative,beaulieu2025observation,open_meta1,open_meta2}. 
Harnessing and accelerating such relaxation processes is therefore both challenging and of timely importance for practical applications of quantum technologies: it may expedite the creation of resource states for quantum enhanced metrology \cite{RevModPhys.89.035002,giovannetti2004quantum,pirandola2018advances}, may be employed to speed up quantum computation \cite{verstraete2009quantum,cirac2012goals,lloyd1996universal} and can be utilized to devise rapid search algorithms \cite{wei2016duality,PhysRevLett.117.150501}. 
One strategy to accomplish such acceleration is to tune system parameters and/or the coupling between the system and its environment \cite{diehl2008quantum,orzel2001squeezed,mi2024stable,PhysRevLett.110.120402,PhysRevLett.119.150502,lin2013dissipative,liu2025general,ulvcakar2025conserved}, in order to increase relaxation rates. 
Another strategy, which is related to the so-called Mpemba effect \cite{mpemba1969cool,lu2017nonequilibrium,MPEgranularfluids,baity2019mpemba}, exploits that the relaxation rate may depend on the initial state of the many-body dynamics. Applying a fine-tuned unitary operation on the initial state may then partially or fully eliminate the excitation of slowly decaying modes and thereby allows to achieve a significant relaxation speedup \cite{nava_MPE,carollo2021mpemba,TH_MPE,ares2025quantum,TH_MPE,MPE_dot_reservoirs,Kochsiek_mpe,beato2025relaxation,westhoff2025fast}.

\begin{figure}
    \centering    \includegraphics[width=\linewidth]{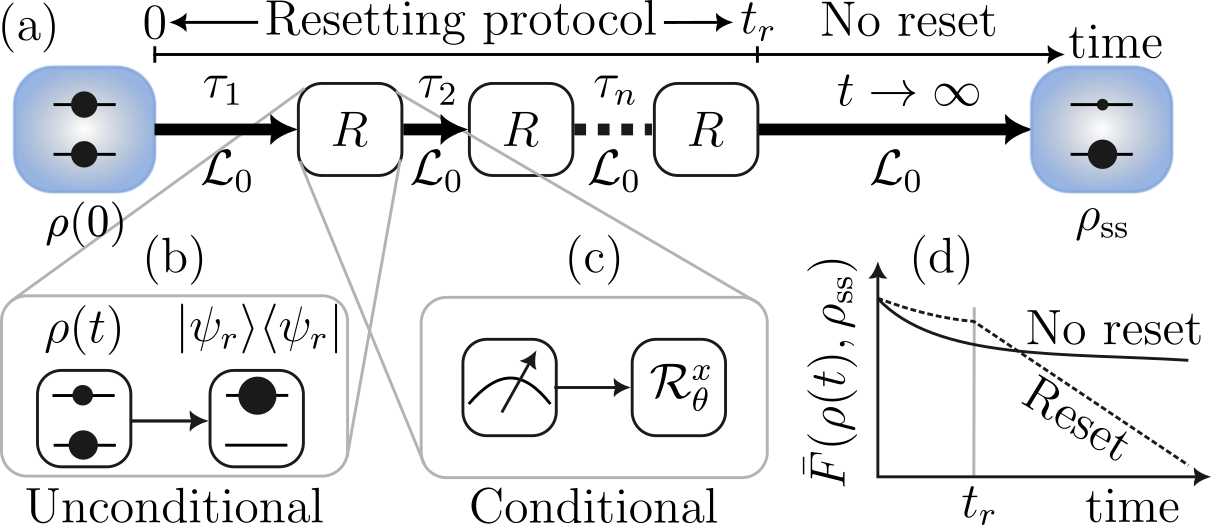}   \caption{\textbf{Illustration of stochastic resetting protocols and relaxation speedup.} (a) For simplicity, we showcase a single qubit system, initialized in the state $\rho(0)$ and evolving under a reset-free Liouvillian $\mathcal{L}_0$. Times $\tau_j$ elapsing between consecutive reset projections, sketched with $R$ boxes, are random. Stochastic resets take place up to an initial transient time $t_r$: $0< t \leq t_r$. After this time, $t>t_r$, the system relaxes to the steady state $\rho_\text{ss}$ of $\mathcal{L}_0$. We consider two resetting protocols: $(b)$ unconditional, where the system resets to a fixed state $\ket{\psi_r}$ regardless of the instantaneous state $\rho(t)$, and $(c)$ conditional, where a state-specific operation (a rotation $\mathcal{R}_\theta^x$ gate in the qubit case) is applied based on measurement outcomes (measurement symbol). $(d)$ Both protocols lead to universal acceleration of relaxation to the stationary state. This is quantified by the faster decay in time of the distance measure $\bar{F}(\rho(t),\rho_\text{ss})$ to zero.}
\label{fig:fig1}
\end{figure}

In this work, we investigate a protocol, illustrated in Fig.~\ref{fig:fig1}(a), that achieves relaxation speedup by superimposing stochastic resetting onto the quantum dynamics, during an initial transient time of finite duration $t_r$. Resets return the system to a designated state \cite{evans2020review,nagar2023review,Sandev_book}, which is a mechanism that has found applications in accelerating search processes in both classical \cite{RevModPhys.83.81,evans2011resetting,evans2013optimal,firstpassageLevi1,firstpassageLevi2} and quantum settings \cite{grunbaum2013recurrence,Barkai2017,Barkai2018,tornow2023,majumdar2023,Walter25,yin2025restart,restart_Yin,Modak_resetting,roy2025causality}. In the case we consider here, transient resetting drives the system toward a non-equilibrium state, which, when used as the initial state of the subsequent reset-free dynamics, accelerates the approach to stationarity. \textcolor{black}{Aspects of this approach have been demonstrated in classical systems \cite{busiello2021inducing}} and in small quantum systems \cite{bao2022acceleratingrelaxationmarkovianopen}, but many pressing questions remain unexplored. These include the universality of the protocol, its scalability to genuine many-body effects such as phase transitions, and the impact of more advanced reset protocols involving conditional nonunitary operations.

To address these questions, we begin by examining the \textit{unconditional} resetting protocol, cf. Fig.~\ref{fig:fig1}(b), where the system is reset to a prescribed state \cite{Rose2018spectral,perfetto2021designing,Magoni_PRA_quantum,dattagupta2022stochastic,sevilla2023dynamics,kulkarni2025dynamically,wald2025stochastic,soldner2025nonanaliticities}. We demonstrate universality of relaxation speedup via resetting, 
by first considering the exemplary case of the driven-dissipative Kerr oscillator \cite{minganti2018dissipative, casteels2017critical, drummond1980quantum,hofheinz2008generation,generation_nonclassical,preparation_fock_states}. The latter features a first-order phase transition with diverging relaxation times. \textcolor{black}{Resetting removes slow metastable relaxation, enabling the rapid preparation of stationary states, which are essential for improved quantum sensing and metrology \cite{sensing_1,sensing_2,sensing_3,sensing_4,sensing_5}}. 
The choice of the initial and reset states in such protocols does not fundamentally require detailed knowledge of the eigenmodes of the dynamics. 
Only macroscopic features of the target stationary states are necessary, such as 
the stationary value of the order parameter.
To completely eliminate the initial state dependence, we propose a second protocol termed \textit{conditional} resetting, sketched in Fig.~\ref{fig:fig1}(c). In this approach, the choice of the reset state is conditional on the measurement output. 
Remarkably, this protocol achieves speedup of relaxation both for few-body systems and many-body ones, such as the Dicke model \cite{PhysRev.93.99,carollo2021mpemba,PhysRevA.99.033845,imai2019dynamical} and the XXZ spin chain, for \textit{any} choice of the initial state. This is in stark contrast with earlier studies, where fine-tuned manipulations of initial states are required \cite{carollo2021mpemba,TH_MPE,MPE_dot_reservoirs,Kochsiek_mpe,beato2025relaxation}. 
\textcolor{black}{Our results, therefore, not only corroborate the understanding of relaxation processes in quantum dynamics, but also open a general pathway for controlled relaxation, enhanced quantum sensing, and accelerated state preparation \cite{westhoff2025fast,zhan2025rapid,liu2025general,chatterjee2025directexperimentalobservationquantum,bagui2025detectionmpembaeffectgood} in many-body open quantum systems.}

\paragraph{Resetting protocol.---}
We consider an open quantum dynamics governed by the Markovian master equation \cite{breuer2002,gorini1976,lindblad1976} ($\hbar=1$ henceforth) with Liouvillian $\mathcal{L}_0$:
\begin{equation}
    \dot{\rho} = \mathcal{L}_0 (\rho)=-i[H,\rho]+\sum_{\mu} D[\mathcal{O}_{\mu}],
\label{eq:master_eq}
\end{equation}
where $H$ is the system Hamiltonian, while $D[\mathcal{O}_{\mu}]= \mathcal{O}_{\mu}\rho\mathcal{O}_{\mu}^\dagger-\{\mathcal{O}_\mu^\dagger\mathcal{O}_\mu,\rho\}/2$ is the dissipator associated to the jump operator $\mathcal{O}_{\mu}$.
The time evolved density matrix $\rho(t)$ can be expressed in terms of the right (left) $R_k(L_k)$ eigenmatrices of $\mathcal{L}_0$ and the corresponding eigenvalues $\lambda_k$ as
\begin{align}   \rho(t)=\rho_{\text{ss}}+\sum_{k\geq2}e^{\lambda_k t}a_k R_k.
\label{eq:decomposition_reset_free}
\end{align}
Here, the coefficients $a_k=\mathrm{Tr}[L_k^\dagger \rho(0)]$, $R_k(L_k)$ contain the information about the initial state $\rho(0)$ via the overlap of the latter with the $k$-th eigenmode. The left and right eigenmatrices satisfy the orthonormality condition $\mathrm{Tr}[R_{k}]=\delta_{k1}$ and $\mathrm{Tr}[L_k^\dagger R_j]=\delta_{k,j}$.
The steady state $\rho_{\text{ss}}(=R_1)$ corresponds to the eigenvalue $\lambda_1=0$ and the relaxation rate is set by the inverse spectral gap, which is defined in terms of the first nonvanishing eigenvalue $Re[\lambda_2]\neq 0$.

It is thus necessary to reduce the contribution from the slowest modes $k=2,3,\dots$ in order to accelerate the relaxation of the system. Here, we achieve this by monitoring the dissipative dynamics via resets and measurements up to a transient time $t_r$, as shown in Fig.~\ref{fig:fig1}(a). The ensuing dynamics $\dot{\rho}=\mathcal{L}_r(\rho)$ is ruled by a modified Liouvillian $\mathcal{L}_r$. After time $t_r$, the resetting channel is switched off so that the system evolves under the reset-free generator $\mathcal{L}_0$. 
We consider two different resetting protocols.

We first focus on the \textit{unconditional resetting} protocol, sketched in Fig.~\ref{fig:fig1}(b). Here, the dissipative dynamics in Eq.~\eqref{eq:master_eq} is interspersed with resets. When a reset takes place, the state of the system is projected to a predefined reset state $\ket{\psi_r}$ \footnote{The reset state does not need to be pure. In the Supplemental material, we also consider examples with a mixed reset state.}. The waiting times between consecutive resets are random variables. When they are distributed according to a Poissonian distribution, this amounts to resetting at constant rate $\Gamma$, which does not depend on the instantaneous state of the system. The ensuing dynamics is then markovian \cite{Hartmann2006,Linden2010,Rose2018spectral,Mukherjee2018,quantum_reset_ldev} and dictated by the Liouvillian $\mathcal{L}_r$: 
\begin{equation}
\mathcal{L}_r(\rho)=\mathcal{L}_0(\rho)+\Gamma\, \mathrm{Tr}(\rho) 
\ket{\psi_r}\!\bra{\psi_r}-\Gamma  \rho.
\end{equation}
The second term on the r.h.s. describes the incoming probability flow towards the reset state $\ket{\psi_r}$ from all the other states, while the third term describes the escape rate from any state due to resetting.

The second protocol we consider is \textit{conditional} resetting. The crucial difference compared to the previous protocol is that here the reset state is not fixed, but it is chosen within a set of $r=1,2,\dots \mathcal{R}$ reset states conditionally on the output of a measurement, cf. Fig.~\ref{fig:fig1}(c). 
The Liouvillian $\mathcal{L}^c_r$ describing conditional resetting is
\begin{equation}    \mathcal{L}_r^c(\rho)=\mathcal{L}_0(\rho)+\Gamma\sum_{r=1}^{\mathcal{R}} \mbox{Tr}[\rho(t) P_r]\ket{\psi_r}\bra{\psi_r}-\Gamma \rho,
\label{eq:conditional_meq}
\end{equation}
where $P_r=\sum_{\mu}\ket{\phi_{\mu}^r}\bra{\phi_{\mu}^r}$ is a projection operator over the eigenspace $\{\ket{\phi_{\mu}^r}\}$ spanned by the measurement outcome. Completeness of the measurement enforces that $\sum_{r=1}^{\mathcal{R}}P_r=\mathbb{I}$. From the second term on the right-hand side, it is evident that resetting to the state $\ket{\psi_r}$ takes place with an effective rate $\sim \Gamma \,\mbox{Tr}[\rho(t) P_r]$ determined by the instantaneous state of the system.
Conditional resetting protocols have been recently implemented \cite{rqh3,rqh4} using dynamic circuits on superconducting qubits, where conditional bit-flip gates are performed based on the outcome of a mid-circuit measurement. 

In both resetting protocols, the state $\rho_r(t)$ in the presence of resetting can be expressed in terms of the eigenvalues and eigenmatrices of the reset-free Lindbladian $\mathcal{L}_0$ such that 
\begin{equation}
\rho_r(t\leq t_r)=\rho_{\text{ss}}+\sum_{k\geq 2}a_k^r(t)e^{\lambda_k t}R_k. 
\label{eq:resetting_fundamental_1}
\end{equation}
For $t>t_r$, the system evolves under the reset-free dynamics as follows 
\begin{equation}
\rho_r(t>t_r)=\rho_{\text{ss}}+\sum_{k\geq 2}a_k^r(t_r)e^{\lambda_k t}R_k, 
\label{eq:resetting_fundamental_2}
\end{equation}
where $a_k^r(t_r)$ are the modified overlaps with the $k$-th eigenmode. 
We quantify relaxation to the stationary state using the measure $\bar{F}(\rho(t),\rho_{\text{ss}})=1-F(\rho(t),\rho_{\text{ss}})$
where $F(\rho(t),\rho_{\text{ss}})$ 
is the fidelity \cite{SM} between the steady state ($\rho_{\text{ss}}$) and the state ($\rho(t)$) at time $t$.
The distance measure $\bar{F}$ vanishes in the long-time limit as the stationary state $\rho_{\mathrm{ss}}$ is approached. A faster relaxation of this decay can be achieved in two ways. First, the weak Mpemba effect (WME) occurs when the absolute value of the modified overlap $|a^r_k(t_r)|$ is reduced compared to the reset-free dynamics \textcolor{black}{for $k\geq 2$}, but not canceled. Second, the strong Mpemba effect (SME) involves the more stringent condition where the modified overlap $|a_k^r(t_r)|=0$ can be exactly nullified \textcolor{black}{for $k\geq 2$}, as shown in Fig.~\ref{fig:fig1}(d). We begin analyzing these phenomena in the simpler protocol of unconditional resetting.

\begin{figure}
    \centering
    \includegraphics[width=\linewidth]{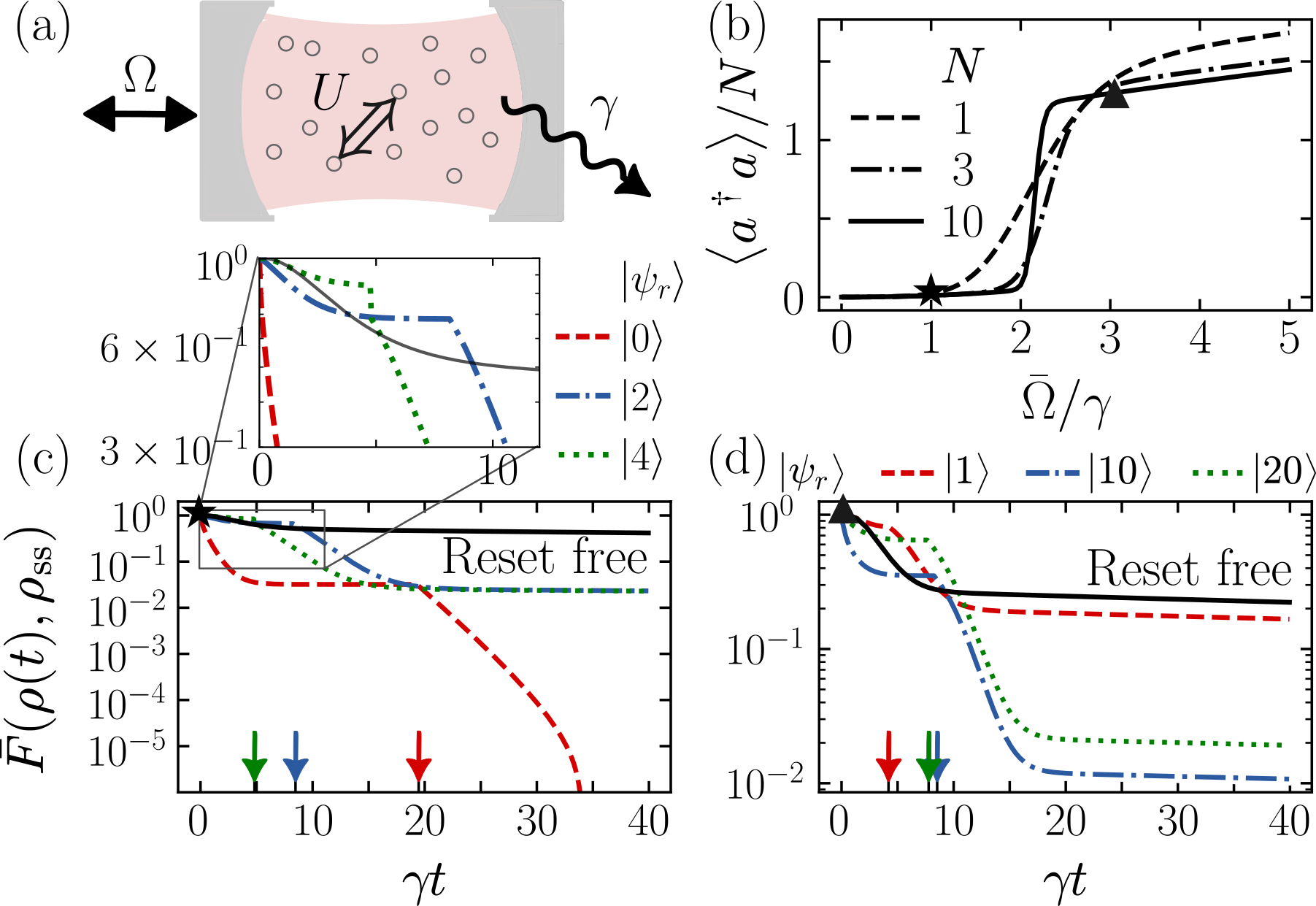}
    \caption{\textbf{Mpemba effects across 1st-order transition. } (a) Sketch of driven dissipative Kerr oscillator, characterized by a drive strength $\Omega$, Kerr nonlinearity $U$, and a dissipation rate $\kappa$. (b) This model exhibits a 1st-order dissipative phase transition \textcolor{black}{in the thermodynamic limit ($N\rightarrow \infty$) of large occupation of the cavity mode}. In panel (c), we analyze the case $\bar{\Omega}/\gamma=1$ with $\bar{\Omega}=\Omega/\sqrt{N}$, where the system relaxes to a \textcolor{black}{vacuum state}. One finds SME for the reset state $\ket{\psi_r}=\ket{0}$, while a WME emerges for other reset states. The initial state is taken as $\rho(0)=\ket{10}\bra{10}$. In panel (d), we consider the case of $\bar{\Omega}/\gamma=3$, where a nonvanishing stationary occupation of the cavity is obtained. A WME is observed for any reset state with a nonzero Fock occupation. The initial state is $\rho(0)=\ket{0}\bra{0}$. In (c) and (d), solid lines indicate dynamics without resets, while dashed and dashed-dotted lines represent dynamics with resets to different states $\ket{\psi_r}$ (see labels) up to time $t_r$ (indicated by arrows). For $t>t_r$, the system follows the reset-free dynamics. Parameters are fixed as $N=10$ and  $\Delta/\gamma=\bar{U}/\gamma=10$ where $\bar{U}=UN$. 
     }
    \label{fig:fig3}
\end{figure}

\paragraph{Mpemba effect across 1st-order phase transition.---} \textcolor{black}{Metastable slow-relaxation is an inherent property of many-body systems featuring a first-order phase transition. Here, we show that the \textcolor{black}{unconditional resetting protocol allows for accelerating relaxation in this challenging dissipative phase transition scenario.} We do this by taking the driven-dissipative Kerr oscillator as an example, see Fig.~\ref{fig:fig3}(a).}
The cavity mode $a$ is here driven by a laser field with strength $\Omega$, while $\Delta$ is the detuning between the drive and the cavity. The radiation mode is self-interacting with strength $U$ and the Hamiltonian is $H=-\Delta a^\dagger a+(U/2)\,a^\dagger a^\dagger a a+\Omega(a+a^\dagger)$. Dissipation of photons from the cavity at rate $\gamma/2$ is modeled via the jump operator $\mathcal{O}_1=\sqrt{(\gamma/2)}a$.
\textcolor{black}{We define $\Omega=\bar{\Omega} \sqrt{N}$ and $U=\bar{U}/N$ where $N$ in a dimensionless parameter proportional to the occupation number of the cavity mode. The thermodynamic limit is obtained for $N\rightarrow \infty$, while keeping $U\vert\Omega\vert^2$ fixed, such that the number of excitations diverges \cite{casteels2017critical}.}
The first-order phase transition and metastability are emergent phenomena since the spectral gap given by $\lambda_2$ vanishes in the thermodynamic limit \cite{minganti2018dissipative,casteels2017critical,drummond1980quantum,open_meta1,open_meta2}. The associated order parameter is given by the cavity occupation density $\braket{a^{\dagger}a}/N$, which discontinuously jumps from zero to a finite value at a specific value of $\bar{\Omega}/\gamma = \bar{\Omega}_c/\gamma$, as illustrated in Fig.~\ref{fig:fig3}(b).

We first study the Mpemba effect for the case $\bar{\Omega}/\gamma < \bar{\Omega}_c/\gamma$, where the cavity has vanishing stationary occupation.
The initial $\rho(0)$ and reset state $\ket{\psi_r}$ are Fock states $\ket{n}$. 
Remarkably, relaxation speedup is achieved for any choice of both the states, cf. Fig.~\ref{fig:fig3}(c). For the fine-tuned choice of $\ket{\psi_r}=\ket{0}$, one further has SME and exponential acceleration. 
For $\bar{\Omega}/\gamma > \bar{\Omega}_c/\gamma$, the system exhibits finite occupation in the steady state. In this case, cf. Fig.~\ref{fig:fig3}(d), WME is observed. The only requirement is that the reset state $\ket{\psi_r}$ presents a nonvanishing, but otherwise arbitrary, cavity occupation. In both phases, we therefore see that neither the initial nor the reset state requires fine-tuning. The latter is fixed solely on the basis of the knowledge of the order parameter of the phase transition.
In addition, we observe that at time $t_r$ the state resulting from resetting is further from stationarity than the one obtained in the reset-free case.
However, for $t>t_r$, the dynamics relaxes more rapidly toward the steady state than in the no-reset case. This behavior is the analog in open quantum systems of the conventional Mpemba effect \cite{mpemba1969cool,lu2017nonequilibrium,MPEgranularfluids,nava_MPE,baity2019mpemba,carollo2021mpemba,ares2025quantum}. 
\textcolor{black}{In the Supplemental Material \cite{SM}, we further discuss the Mpemba effect via unconditional resetting in few-level systems, while we now move to central protocol developed in this manuscript, which is conditional resetting.}

\paragraph{Conditional reset in the Dicke model.---}  \textcolor{black}{The Mpemba effect via the unconditional resetting protocol depends on both the initial and reset states. Nevertheless, in many cases, e.g., in quantum computation \cite{nielsen2010quantum}, neither of these can be prepared precisely due to noise.} This problem can be (partially) overcome by exploiting conditional resetting in Eq.~\eqref{eq:conditional_meq}. 
\textcolor{black}{To showcase the universality of the conditional resetting protocol in the many-body realm, we explore the open Dicke model in this section, a driven XXZ spin chain in the next section, \textcolor{black}{while the Kerr oscillator is presented in the End Matter.}} 

\textcolor{black}{The spin-only representation of the open Dicke model is governed by the following master equation \cite{PhysRevA.99.033845,imai2019dynamical,SM}}
\begin{equation}
\dot{\rho}=\mathcal{L}_0(\rho)=-i[\omega \hat{S}_z -\frac{g_{\text{eff}}\omega_c}{N}\hat{S}^2_x,\rho ]+\frac{g_{\text{eff}}\kappa}{N} D[\hat{S}_x]\rho. \label{eq:dicke}
\end{equation}
\textcolor{black}{where $g_{\text{eff}}=4 g^2/(4\omega_c^2+\kappa^2)$.
Here, $\hat{S}_\alpha=\sum_{k=1}^N\sigma_\alpha^k$ are the collective spin operators, with $\sigma_\alpha^k$ being the Pauli spin-operator of the $k$-th atom and $\alpha\in \{x,y,z\}$. 
The Hilbert space of the collective spin system is spanned by the $N+1$ dimensional Dicke eigenbasis $\vert S, m_z\rangle $, where $S=N/2$ is the total spin length and $m_z\in\{S, S-1,\ldots,-S\}$ is the magnetization along the $z$ direction.
The stationary state of Eq.~\eqref{eq:dicke} is fully mixed ($\mathbb{I}/(N+1)$) and has zero magnetization in the $z$ direction, since the jump operator is hermitian \cite{PhysRevA.99.033845,imai2019dynamical}.
}

\textcolor{black}{We consider a conditional resetting protocol where a measurement of the magnetization in the $z$ direction is performed prior to resetting. We design conditional resetting to act as a ``magnetization thermostat'': if the measured value is positive (negative) $m_z>0$ $(m_z<0)$, the system is reset to $\vert S,-S \rangle$ ($\vert S,S \rangle$). In the case zero magnetization $m_z=0$ is measured, the system is reset to either $\vert S,S \rangle$ or $\vert S,-S \rangle$, the two states chosen with equal probability. This conditional resetting protocol is implemented by the jump operators}
\begin{align*}
    L^{-S}_k&=\sqrt{\Gamma} \vert S,-S\rangle \langle S,m_z\vert, ~\forall m_z\in \{S,\ldots,1\}\nonumber,\\ 
    L^{S}_k&=\sqrt{\Gamma} \vert S,S\rangle \langle S,m_z\vert, ~\forall m_z\in \{-1,\ldots,-S\}, \nonumber\\
    L^{-S}_0&=\sqrt{\Gamma/2} \vert S,-S\rangle \langle S,0\vert,
    L^{S}_0=\sqrt{\Gamma/2} \vert S,S\rangle \langle S,0\vert.    
\end{align*}
\begin{figure}[t!]
    \centering
\includegraphics[width=\linewidth]{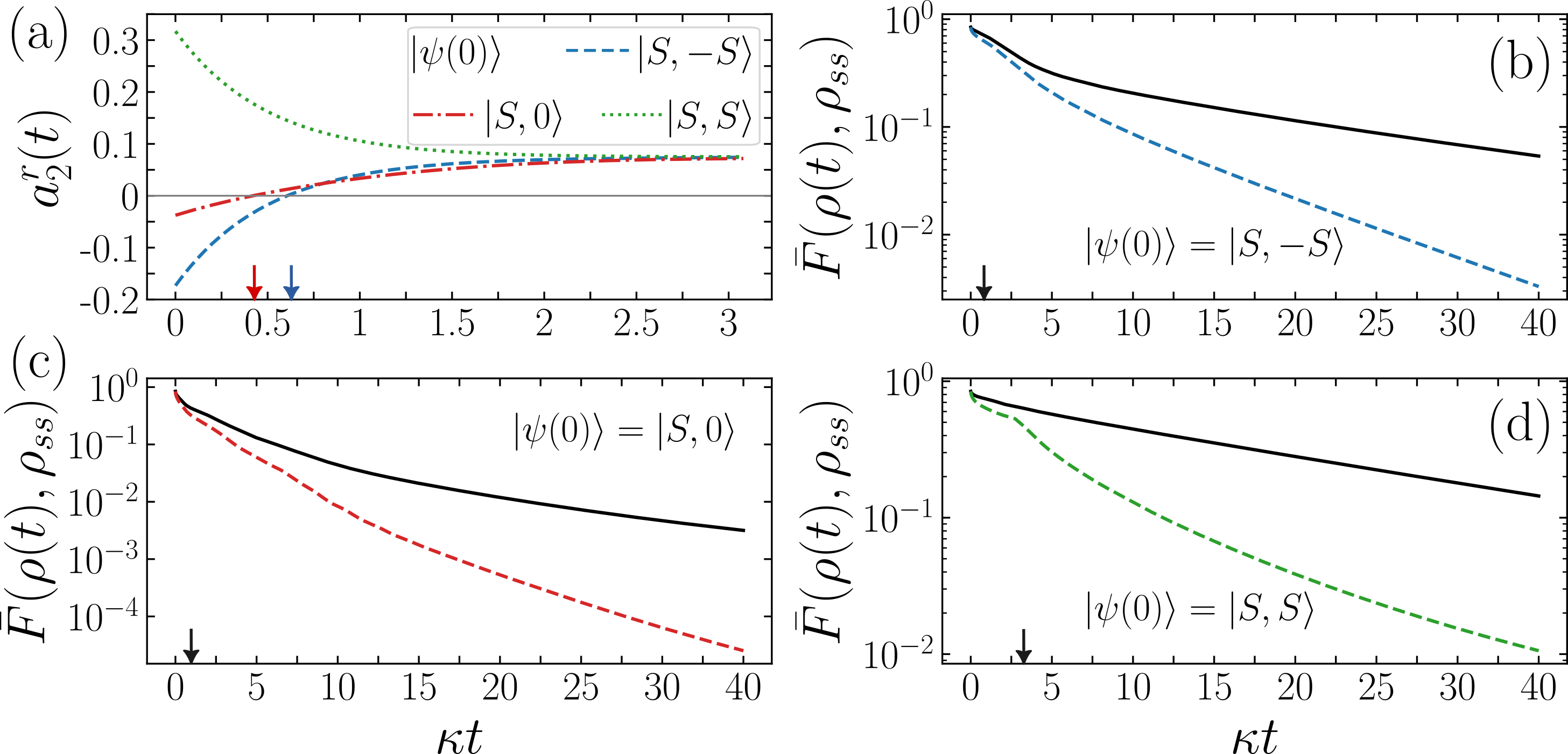}
    \caption{\textbf{Mpemba effect in Dicke model.} \textcolor{black}{(a) Coefficient $a_2^r(t)$ for the three initial states reported in the legend. In panels (b), (c) and (d), we report the dynamics of the fidelity from the initial states $\vert S,-S\rangle $, $\vert S,0\rangle $ and $\vert S,S \rangle $ used in (a), respectively. The arrows mark the time $t_r$ when resetting is stopped. Solid line represents the reset-free dynamics, while the dashed line is the dynamics with resetting up to time $t_r$. Panels (b) and (c) show a strong Mpemba effect for initial states $\vert S,-S\rangle $ and $\vert S,0\rangle $, respectively, while panel (d) indicates a weak Mpemba effect for initial state  $\vert S,S \rangle $. Parameter values are: $\omega_c/\omega=\kappa/\omega=g/\omega=\Gamma/\omega=1$ and $N=20$. }}
    \label{fig:fig5}
\end{figure}


\textcolor{black}{It induces both SME and WME in the Dicke model, as shown in Fig.~\ref{fig:fig5}. SME occurs for initial states where  $a_2^r(0)<0$ in Eqs.~\eqref{eq:resetting_fundamental_1} and \eqref{eq:resetting_fundamental_2}, since the given resetting protocol results in a positive value of $a_2^r(t)$ for long times $t$ irrespective of the chosen initial state, see Fig.~\ref{fig:fig5}(a).
Consequently, all initial states with negative $a_2^r(0)$ will eventually cross zero.
Once this has happened (see vertical arrows in  Fig.~\ref{fig:fig5}(a)), the protocol can be stopped, leading to SME as shown in Figs.~\ref{fig:fig5}(b,c). Also for initial states with $a_2^r(0)>0$, this protocol reduces the overlap with the second eigenmode over time, as shown in Fig.~\ref{fig:fig5}(a), resulting in WME, see Fig.~\ref{fig:fig5}(d). We therefore see that conditional resetting induces universal relaxation speedup, regardless of the chosen initial state, solely based on macroscopic information about the targeted stationary state. 
} 


\paragraph{Conditional reset in the driven XXZ spin chain.---}
\textcolor{black}{In order to showcase the generality of our analysis beyond collective spin models, we here consider the XXZ spin chain with a transverse field $\Omega$ (open boundary conditions are henceforth assumed) with local dephasing so that the dynamics is described by the following master equation} 
\begin{align}
    \mathcal{L}_0(\rho)=-i[H_{\text{XXZ}}+\Omega\sum_{i=1}^{N} \sigma_i^x,\rho]+\gamma \sum_{i=1}^{N} D[\sigma^i_z]\rho,
\end{align}
\textcolor{black}{where $H_{\text{XXZ}}=\sum_{i=1}^{N-1}[J_{xy}(\sigma^x_i\sigma^x_{i+1}+\sigma^y_i\sigma^y_{i+1})+J_z\sigma^z_i\sigma^z_{i+1}]$. 
In the absence of a magnetic field ($\Omega=0$), the system exhibits $U(1)$ symmetry, conserving the $z$ component of the magnetization, $\hat{S}_z$. Consequently, the maximally mixed state within each fixed-magnetization sector is stationary. To avoid this fine-tuned situation, we consider $\Omega\neq0$, which breaks the $U(1)$ symmetry and yields a unique, nontrivial many-body stationary state.
}

\textcolor{black}{
Since the stationary state for $\Omega \neq 0$ is mixed and spans all magnetization sectors, we introduce a resetting protocol similar to the Dicke model that acts as a ``magnetization thermostat" to enhance population across all sectors. Specifically, we measure the magnetization $\hat{S}_z$: if $m_z>0$, the system is reset to $\ket{\psi_r^{\downarrow}}=\ket{\downarrow\downarrow\cdots\downarrow}$; if $m_z<0$, it is reset to $\ket{\psi_r^{\uparrow}}=\ket{\uparrow\uparrow\cdots\uparrow}$; and for $m_z=0$, either state is chosen with equal probability. This protocol is implemented through the following jump operators:
}
\begin{align*}
L_k^{s_z=0}&=\sqrt{\Gamma/2}\ket{\psi_r^{\downarrow}}\bra{\phi_k^{s_z=0}},  L_k^{s_z=0}=\sqrt{\Gamma/2}\ket{\psi_r^{\uparrow}}\bra{\phi_k^{s_z=0}},\nonumber \\
       L_k^{s_z>0}&=\sqrt{\Gamma}\ket{\psi_r^{\downarrow}}\bra{\phi_k^{s_z>0}}, L_k^{s_z<0}=\sqrt{\Gamma}\ket{\psi_r^{\uparrow}}\bra{\phi_k^{s_z<0}}.
\end{align*}
\textcolor{black}{Here, $\ket{\phi_k^{s_z>0}},\ket{\phi_k^{s_z>0}}$ and $\ket{\phi_k^{s_z=0}}$ represents the element of the sets $\phi^{s_z>0}$, $\phi^{s_z<0}$ and $\phi^{s_z=0}$ of product states in the $z$ direction with $m_z>0$, $m_z<0$ and $m_z=0$, respectively. The union of these three sets forms a complete eigenbasis of the Hilbert space, with their elements obeying the orthonormality condition $\langle \phi_k^\mu \vert \phi_{k^{'}}^{\mu^{'}} \rangle=\delta_{k,k^{'}}\delta_{\mu,\mu^{'}}$.}

\begin{figure}[t]
    \centering
    \includegraphics[width=\linewidth]{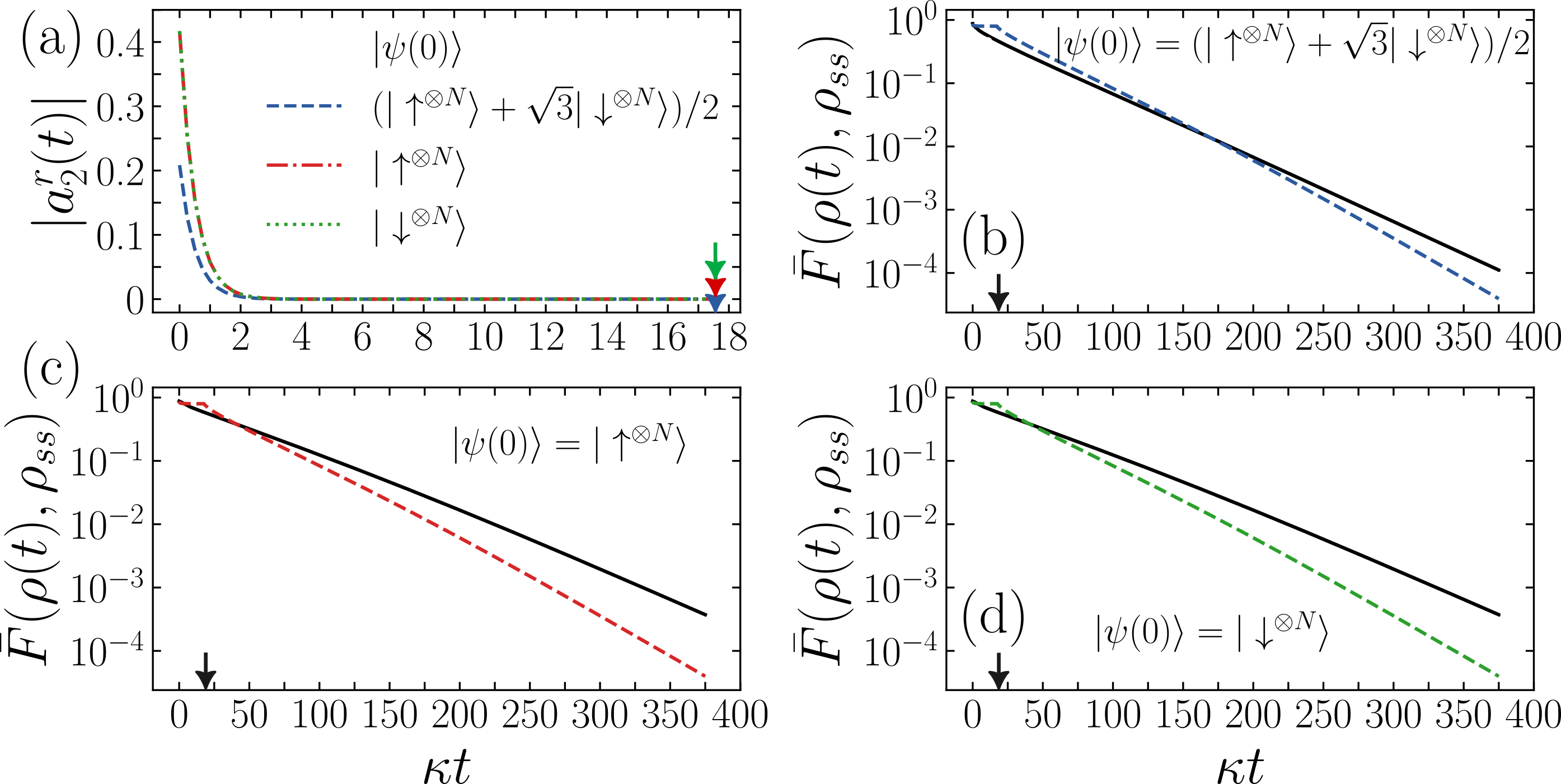}
    \caption{\textbf{Mpemba effect for the XXZ spin chain in a magnetic field.} \textcolor{black}{In panel (a), we show that the coefficient \(a_2^r(t)\) converges to zero for all initial states reported. This results in exponential speedup of relaxation --- strong Mpemba effect. Panels (b), (c), and (d) illustrate the dynamics of the fidelity for the three initial states considered in panel (a). Solid lines correspond to the no-reset case, while dashed lines to the result with transient resetting up to time $t_r$ (indicated by the arrows). Parameter values are: $N=6$, $J_{xy}/\gamma=1/4$, $J_z/\gamma=1$, $\Omega/\gamma=0.1$ and $\Gamma/\gamma=1$.}} 
    \label{fig:fig6}
\end{figure}

\textcolor{black}{In Fig.~\ref{fig:fig6}, we show that this protocol induces the SME regardless of the initial state. In particular, the coefficient $a_2^r(t)$ is exactly canceled at $t_r$, irrespective of the initial state chosen, as shown in Fig.~\ref{fig:fig6}(a). Interestingly, one observes in all the cases in Figs.~\ref{fig:fig6}(b,c,d) that the many-body state at the time $t_r$, when resetting is switched off, is farther from the steady state than the no-reset cases. This is quantified by the corresponding values for the fidelity $\bar{F}(\rho(t_r),\rho_{\text{ss}})$. In spite of this, for $t>t_r$, the curves for the fidelity with transient resetting (dashed lines) cross with those obtained without resetting (solid lines) and one has faster convergence to stationarity. This phenomenon is analogous to the conventional Mpemba effect. With this paradigmatic example, we have thus provided evidence that conditional resetting leads to universal acceleration of relaxation, even in the generic case of a many-body open quantum system.}

\paragraph{Conclusion.---} 
We demonstrated that introducing \textcolor{black}{transient resetting} results in a universal speedup of the relaxation of open many-body quantum systems, eliminating the need for fine-tuning of initial states.
For conditional resetting, acceleration is achieved completely independently of the initial state. \textcolor{black}{Remarkably, the effect persists even for many-body systems exhibiting a first-order dissipative phase transition with emergent long relaxation timescales, and for non-integrable XXZ spin chain in a transverse field.} 
The examples and mechanisms discussed in this work are amenable to experimental implementation in several well-established platforms, including cavity resonators \cite{leghtas2015confining} and superconducting circuits \cite{chen2023quantum,beaulieu2025observation} for nonlinear Kerr oscillators, superconducting qubits \cite{rqh3,rqh4}, and trapped ions \cite{bergquist1986shelving} for qutrit systems. 
In the future, it will be interesting to demonstrate our findings on quantum hardware platforms. \textcolor{black}{Along this direction, our results are crucially robust with respect to imperfections in the preparation of the reset and/or initial state \cite{SM}, which are intrinsic issues to current noisy quantum processors. This analysis could thus pave the way for the development of universal state-preparation and control protocols that are robust to noise.}

\paragraph{Acknowledgments.---}  P.S. acknowledges support from the Alexander von Humboldt Foundation through a Humboldt research fellowship for postdoctoral researchers. I.L. acknowledges support from the QuantERA II program (project CoQuaDis, DFG Grant No. 532763411) that has received funding from the EU H2020 research and innovation program under GA No. 101017733. I.L. also acknowledges funding from the Deutsche Forschungsgemeinschaft (DFG, German Research Foundation) through the Research Unit FOR 5413/1, Grant No.~465199066, from Leverhulme
Trust (Grant No. RPG-2024-112) and from the European Union through the ERC grant OPEN-2QS (Grant No. 101164443). G.P. acknowledges funding from the Swiss National Science Foundation (SNSF) through Grant No.~10005336.

\section{End Matter}

\section{Conditional resetting in driven-dissipative Kerr oscillator \label{sec:conditional_Kerr}}
\textcolor{black}{In the main text, we analyzed the Mpemba effect in a driven dissipative Kerr oscillator under unconditional resetting. Here, we extend the discussion to the case of conditional resetting. The dynamics in the absence of resetting for the driven dissipative Kerr oscillator is governed by the Liouvillian
\begin{align}
\dot{\rho} = -i\left[-\Delta a^\dagger a + \frac{U}{2} a^\dagger a^\dagger a a + \Omega (a + a^\dagger), \rho \right] + \frac{\gamma}{2} D[a]\rho,
\end{align}
with analogous definitions of the parameters as in the main text.}

\begin{figure}[b!]
    \centering
\includegraphics[width=\linewidth]{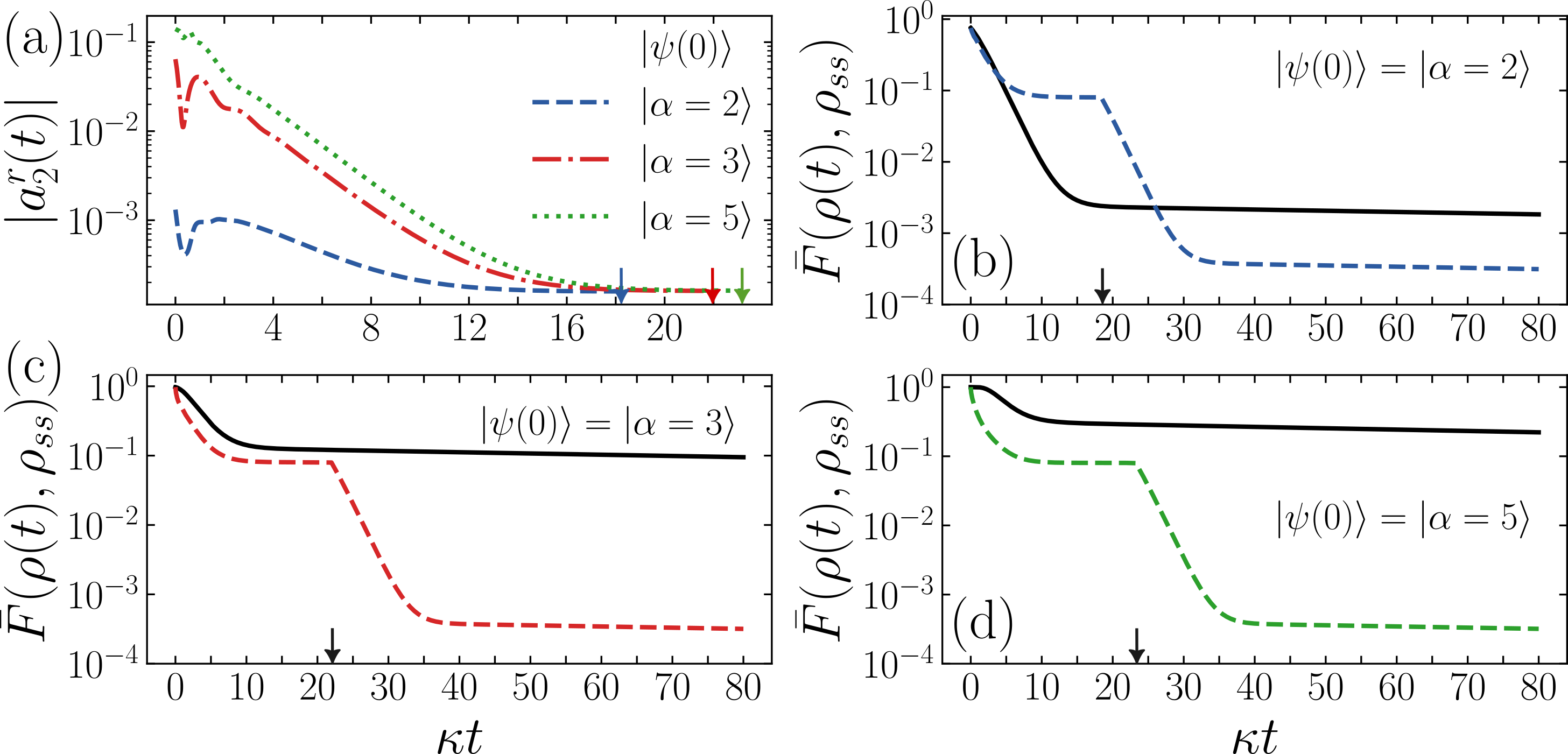}
    \caption{\textcolor{black}{Weak Mpemba effect in the Kerr oscillator through conditional resetting for the following parameter values: $\bar{\Omega}/\gamma=1$, $\Delta/\gamma=\bar{U}/\gamma=N=10$. This configuration leads to a steady state characterized with zero occupation of the cavity. We define the reset state as \(\ket{\alpha=0.5}\) and the cutoff Fock state as $\ket{\bar{n}}=\ket{3}$.
In panel (a), we present the coefficient $|a_2^r(t)|$ for various initial states $\ket{\psi(0)}$ reported in the legend. While $|a_2^r(t)|$ approaches zero, it does not completely vanish. WME is thus induced by conditional resetting. Panels (b), (c), and (d) show the distance measure $\bar{F}(\rho(t),\rho_{\text{ss}})$ for the different initial states in (a). In all the cases, one finds WME and the fidelity is reduced by at least one order of magnitude thereby significantly going beyond the divergent relaxation time scale of the reset-free dynamics (black solid line) due to metastability.}}
    \label{fig:kerr_cond_1}
\end{figure}

\textcolor{black}{We introduce a conditional resetting protocol, in which the system is reset to a coherent state $\ket{\alpha}$ characterized by $\braket{\alpha| a^\dagger a|\alpha } = |\alpha|^2$, only when it is measured in a specific sector of the Hilbert space. We choose to reset to coherent states, motivated by the relative experimental simplicity of preparing coherent states compared to single Fock states \cite{hofheinz2008generation,generation_nonclassical,preparation_fock_states}. The coherent state $\ket{\alpha}$ is defined as 
\begin{equation}
  |\alpha\rangle = e^{-|\alpha|^2/2} \sum_{n=0}^{\infty} \frac{\alpha^n}{\sqrt{n!}} |n\rangle = D(\alpha)|0\rangle,
\label{eq:coherent_state_kerr}   
\end{equation}
where $D(\alpha) = \exp ( \alpha a^\dagger - \alpha^* a)$ is the displacement operator.
We then define a cutoff in the Fock basis, denoted by $\ket{\bar{n}}$. If the system is in the phase with zero (nonzero) occupation of the cavity, then the system is reset to the coherent state $\ket{\alpha}$ in Eq.~\eqref{eq:coherent_state_kerr} only when the measured occupation is above $n>\bar{n}$ (below $n<\bar{n}$); otherwise, the dynamics proceeds from the state the system has been projected on by the measurement. This protocol is solely based on the knowledge of the order parameter of the dissipative phase transition exhibited by the Kerr oscillator and it is implemented through the following set of jump operators:
\begin{align}
    L_k=
\begin{cases}
\ket{\alpha}\bra{k} \quad \mbox{if} \,\, \, k>\bar{n}, \\
    \ket{k}\bra{k} \quad \mbox{if} \,\,\, k\leq \bar{n},
\end{cases}
\label{eq:kerr_conditional_zero}
\end{align}
for the parameter regime where the steady state has zero occupation. For the phase where the steady state has a nonzero occupation, instead, one has the following jump operators
\begin{align}
L_k=
\begin{cases}
    \displaystyle \ket{k}\bra{k} \quad \mbox{if} \,\,\, k>\bar{n}, \\
    \displaystyle \ket{\alpha}\bra{k} \,\,\,\,\, \mbox{if} \,\,\, k\leq \bar{n}.
\end{cases}
\label{eq:kerr_conditional_nonzero}
\end{align}
This protocol thereby corrects the state of the system so as to lower (increase) the occupation of the cavity depending on the phase being described by a vanishing or nonvanishing order parameter, as in Eqs.~\eqref{eq:kerr_conditional_zero} and \eqref{eq:kerr_conditional_nonzero}, respectively.} 

\begin{figure}[h!]
    \centering
    \includegraphics[width=\linewidth]{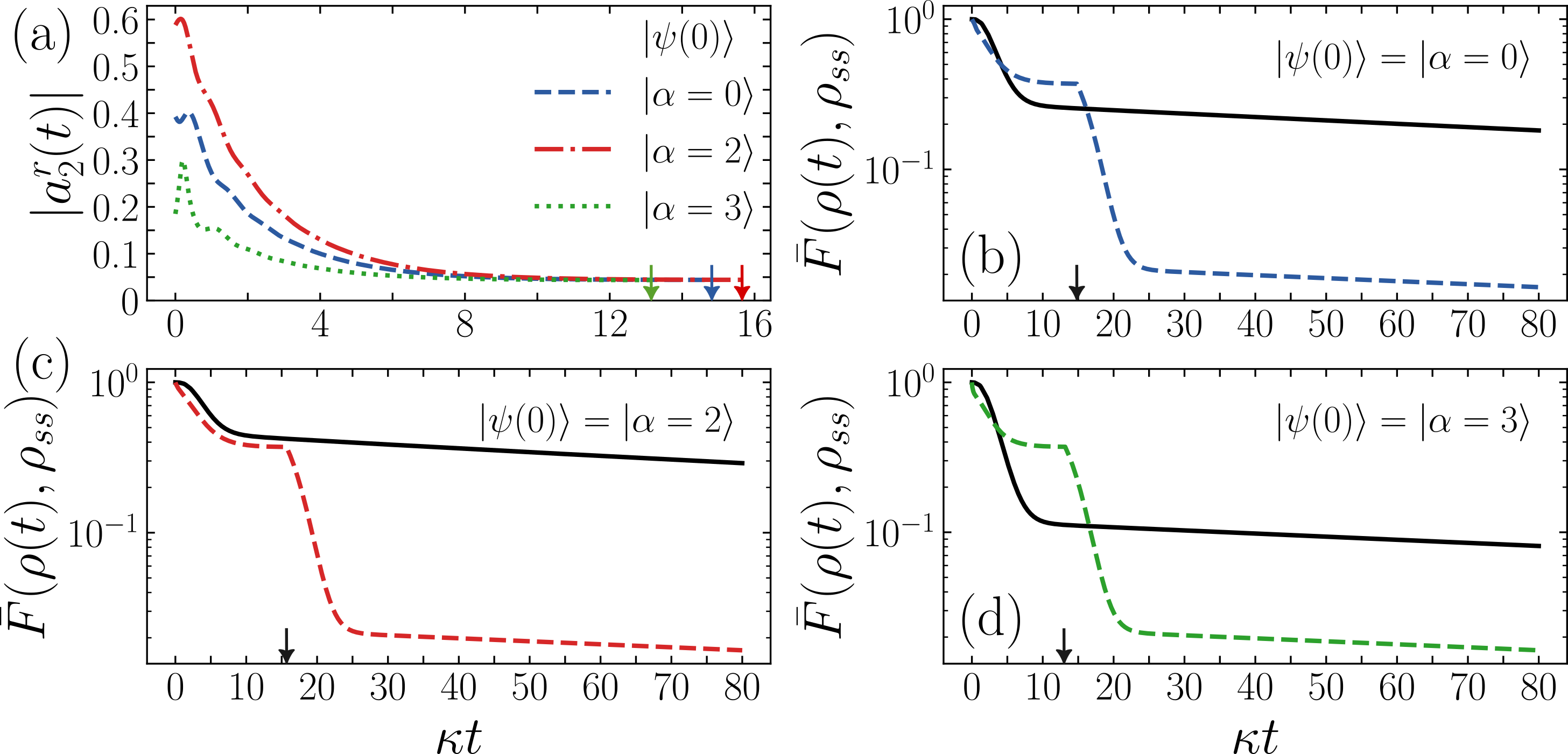}
    \caption{\textcolor{black}{Weak Mpemba effect in the Kerr oscillator through conditional resetting for the parameter values $\bar{\Omega}/\gamma=3,\Delta/\gamma=\bar{U}/\gamma=N=10$, leading to a steady state with finite occupation of the cavity. We consider the reset state to be $\ket{\alpha=3.5}$ and the cutoff Fock state is $\ket{4}$. In panel (a), we plot the coefficient $\vert a_2^r(t)\vert $ for different initial states $\ket{\psi(0)}$ reported in the legend. The distance measure $\bar{F}(\rho(t),\rho_{\text{ss}})$ for the different initial states of (a) is shown in panels (b,c,d). WME effect is observed for all the initial states with a fidelity lowered by approximately one order of magnitude compared to the metastable reset-free evolution (black solid line).}}
    \label{fig:kerr_cond_2}
\end{figure}

\textcolor{black}{In Fig.~\ref{fig:kerr_cond_1}, we report the results for the protocol in Eq.~\eqref{eq:kerr_conditional_zero}, where the system is in the phase with zero occupation. In Fig.~\ref{fig:kerr_cond_2}, instead, we report our results according to the conditional protocol in Eq.~\eqref{eq:kerr_conditional_nonzero} for the phase with a nonzero occupation of the cavity mode. In both the cases we find a weak Mpemba effect for all the chosen initial states. Relaxation is accelerated by at least one order of magnitude in all the cases. For the stationary state with zero stationary occupation, we even find in Fig.~\ref{fig:kerr_cond_1}(c) and (d) acceleration by two orders of magnitude. These results are particularly important in this model, where the reset-free relaxation timescale is divergent due to metastability concurrent with the first-order phase transition taking place in the thermodynamic limit. Also in this complicated driven-dissipative setup with a first-order phase transition, conditional resetting therefore induces universal relaxation speedup. The design of the resetting protocol fundamentally requires knowledge of only the order parameter of the phase transition. Neither detailed knowledge of the stationary state nor fine-tuned control of the initial state is necessary.}

\newpage
\setcounter{equation}{0}
\setcounter{figure}{0}
\setcounter{table}{0}
\renewcommand{\theequation}{S\arabic{equation}}
\renewcommand{\thefigure}{S\arabic{figure}}

\makeatletter
\renewcommand{\theequation}{S\arabic{figure}}
\renewcommand{\thefigure}{S\arabic{figure}}

\onecolumngrid
\newpage

\setcounter{page}{1}

\setcounter{secnumdepth}{3}
\pagestyle{plain}

\begin{center}
{\Large SUPPLEMENTAL MATERIAL}
\end{center}
\begin{center}
\vspace{0.8cm}
{\Large Universal relaxation speedup \textcolor{black}{in nonequilibrium quantum many-body systems} through transient resetting}
\end{center}
\begin{center}
Parvinder Solanki$^{1}$, Igor Lesanovsky$^{1,2}$, and Gabriele Perfetto$^{3}$
\end{center}
\begin{center}
$^1${\em Institut f\"ur Theoretische Physik, Universit\"at T\"ubingen, Auf der Morgenstelle 14, 72076 T\"ubingen, Germany}\\
$^2${\em School of Physics and Astronomy and Centre for the Mathematics and Theoretical Physics of Quantum Non-Equilibrium Systems, University of Nottingham, Nottingham, NG7 2RD, United Kingdom}\\
$^3${\em Institut für Theoretische Physik, ETH Zürich, Wolfgang-Pauli-Str. 27, 8093 Zürich, Switzerland}
\end{center}
\noindent In this Supplemental Material, we present the calculations leading to the results presented in the main text. In Sec.~\ref{sec:unconditional}, we present the general theory behind the emergence of the weak (WME) and strong Mpemba effect (SME) for a Markovian open-quantum system subject to unconditional resetting. In Sec.~\ref{sec:unconditional_qubit}, we specialize the analysis to a qubit coupled to a thermal bath. In Sec.~\ref{sec:unconditional_3_states}, we consider a qutrit displaying metastable dynamics and we assess the impact of resetting to \textcolor{black}{pure and mixed states. In Sec.~\ref{sec:conditional}, we present the conditional reset protocol applied to qubit and qutrit systems. In Sec.~\ref {sec:cond_vs_uncond}, we compare the conditional and the unconditional resetting protocol for these models. The former removes any dependence of the relaxation speedup on the chosen initial state. In Sec.~\ref{sec:unconditional_Dicke}, we eventually conclude by presenting results for unconditional resetting in the many-body Dicke model, which describes atoms in a cavity interacting with a radiation field. Here, we show that relaxation speedup induced by resetting is robust with respect to phase noise in the preparation of the reset state.}

\section{Unconditional reset protocol}
\label{sec:unconditional}
In this section, we provide a detailed discussion of the unconditional reset protocol. 
We adopt the Lindblad description of unconditional resetting processes in open quantum systems, as explored in \cite{Rose2018spectral,perfetto2021designing,quantum_reset_ldev,bao2022acceleratingrelaxationmarkovianopen}.
Let us consider a Markovian open-quantum system that evolves according to a Liouvillian $\mathcal{L}_0$ in the absence of resetting processes. 
The dynamics of the system is then described by the equation:
\begin{equation}
    \dot{\rho} = \mathcal{L}_0 (\rho)=-i[H,\rho]+\sum_{\mu}D[\mathcal{O}_{\mu}], \quad \mbox{with} \quad D[\mathcal{O}_{\mu}]=
    \sum_{\mu}  \left(\mathcal{O}_{\mu}\rho\mathcal{O}_{\mu}^\dagger-\frac{1}{2}\{\mathcal{O}_\mu^\dagger\mathcal{O}_\mu,\rho\}\right),
\label{eq:reset_free}
\end{equation}
where $\rho$ represents the state of the system and $\mathcal{O}_{\mu}$ are jump operators that describe the system-bath interaction. Under the influence of $\mathcal{L}_0$, the system approaches a steady state, denoted as $\rho_{\text{ss}}$. 

To expedite the relaxation process towards the steady state $\rho_{\text{ss}}$, unconditional resetting is introduced.
The unconditional reset to a particular state $\ket{\psi_r}$ is implemented by including additional dissipative processes described by the jump operators $\mathcal{O}_\mu=\sqrt{\Gamma} \ket{\psi_r}\bra{\phi_\mu}$. Here, $\ket{\phi_\mu}$'s form a complete orthonormal basis of eigenvectors $\mu \in\{0,1,2,\dots,\mbox{dim}(\mathcal{H})-1\}$ of the Hilbert space $\mathcal{H}$ and $\Gamma$ is the resetting rate. This is constant as a function of time, and the ensuing equation is Markovian and described by the generator $\mathcal{L}_r$ as follows
\begin{equation} \dot\rho_r=\mathcal{L}_r(\rho_r)=\mathcal{L}_0(\rho_r)+\Gamma \mbox{Tr}(\rho_r) \ket{\psi_r}\bra{\psi_r}-\Gamma  \rho_r,
\label{eq:reset}
\end{equation}
This equation shows that stochastic resetting can be thought of as a dissipative process in addition to the ones present in the reset-free dynamics of Eq.~\eqref{eq:reset_free}. We consider the application of the resetting dynamics $\mathcal{L}_r$ for a time $t_r$. The state of the system initialized with state $\rho(0)$  at time $t$ is consequently expressed as 
\begin{align}
    \rho_r(t) = \rho_{\text{ss}}^r + \sum_{k \geq 2} e^{\lambda_k^r t} \text{Tr}[(L_k^r)^\dagger \rho(0)] R_k^r, \qquad t<t_r, \label{eq:rho_t_reset}
\end{align}
where $R_k^r$ and $L_k^r$ are the right and left eigenvectors, respectively, and $\lambda_k^r$ are the corresponding eigenvalues of $\mathcal{L}_r$.
Here, $\rho_{\text{ss}}^r=R^r_1$ denotes the steady state of the system in the presence of resetting (we adopt the convention where the stationary state corresponds to the mode $k=1$).

To understand the effect of resetting on the dynamics of the system, we can decompose the time evolution of the system undergoing unconditional resetting in terms of the eigenvalues and eigenvectors of the reset-free Liouvillian, $\mathcal{L}_0$, cf. also Refs.~\cite{busiello2021inducing,bao2022acceleratingrelaxationmarkovianopen}. Let us denote the right(left) eigenvectors and eigenvalues of the reset-free Liouvillian $\mathcal{L}_0$ as $R_k$($L_k$) and $\lambda_k$, respectively. Resetting modifies the steady state $\rho_\text{ss}$ of the unperturbed system $\mathcal{L}_0$ as follows \cite{Rose2018spectral}
\begin{equation}
\rho_{\text{ss}}^r=\rho_{\text{ss}}+\sum_{k\geq2}\frac{\mathrm{Tr}[L_k^\dagger \ket{\psi_r}\bra{\psi_r}]}{\Gamma-\lambda_k} R_k. 
\label{eq:stationary_state_resetting}
\end{equation}
Unconditional resetting does not affect the right eigenvectors of $\mathcal{L}_0$ such that $R^r_k = R_k $ for $k\geq 2$. 
In contrast, the corresponding left eigenvectors $L_k^r$ in the presence of resetting can be expressed as 
\begin{equation}
L_k^r = L_k + \frac{\Gamma \langle \psi_r \vert L_k \vert \psi_r \rangle}{(\lambda_k^* - \Gamma)} I,
\end{equation}
in terms of the reset-free $L_k$ ones. The eigenvalues of $\mathcal{L}_0$, except the zero eigenvalue, experience a constant shift of $-\Gamma$ such that 
\begin{equation}
\lambda_{k\neq 0}^r=\lambda_{k\neq 0}-\Gamma.
\label{eq:unconditional_resetting_shift_eigs}
\end{equation}
Therefore, inserting Eqs.~\eqref{eq:stationary_state_resetting}-\eqref{eq:unconditional_resetting_shift_eigs} into Eq.~(\ref{eq:rho_t_reset}), one obtains 
\begin{align}
    \rho_r(t)&=\rho_{\text{ss}}+\sum_{k\geq 2}\underbrace{\Big[\frac{\Gamma d_k}{\Gamma -\lambda_k}e^{-\lambda_k t}+(a_k-\frac{\Gamma d_k}{\Gamma -\lambda_k})e^{-\Gamma  t}\Big]}_{a_k^r(t)}e^{\lambda_k t}R_k \nonumber \\
    &=\rho_{\text{ss}}+\sum_{k\geq 2}a_k^r(t)e^{\lambda_k t}R_k, \qquad t< t_r. \label{eq:quantum_mpemba}
\end{align}
In the previous equation, we introduced the modified overlap $a_k^r$
\begin{align}
    a_k^r(t)=\frac{\Gamma d_k}{\Gamma -\lambda_k}e^{-\lambda_k t}+\left(a_k -\frac{\Gamma d_k}{\Gamma -\lambda_k}\right)e^{-\Gamma  t}, \label{eq:akr}
\end{align}
and the overlaps $a_k$ and $d_k$, which are given by
\begin{equation}
a_k=\mbox{Tr}[L_k^{\dagger} \rho(0)], \qquad  \mbox{and} \qquad d_k=\mbox{Tr}[L_k^{\dagger}\ket{\psi_r}\bra{\psi_r}].
\label{eq:overlaps}
\end{equation}
The coefficient $a_k$ therefore quantifies the overlap of the $k$-th eigenmode with the initial state $\rho(0)$, while $d_k$ contains the information about the reset state $\ket{\psi_r}$. Equations \eqref{eq:quantum_mpemba} and \eqref{eq:overlaps} form the basis of our analysis for the Mpemba effect under the unconditional resetting protocol. In particular, for $t\geq t_r$ resetting is switched off and the dynamics is ruled by the reset-free Liouvillian according to Eq.~(6) of the main text. Relaxation to the stationary state of $\mathcal{L}_0$ is therefore accelerated by minimizing $|a_k^r(t_r)|$ for the slowest decaying modes $k=2,3\dots$. This can be done by tuning the resetting time $t_r$ and/or the reset $\ket{\psi_r}$ and initial $\rho(0)$ initial states. Exact cancellation of the $k$th eigenmode $|a_k^r(t_r)|=0$ can achieved by choosing a time $t_r$ such that the right hand side of Eq.~\eqref{eq:akr} is zero. This leads to 
\begin{equation}
t_r=\frac{1}{\Gamma-\lambda_k}\ln\left(1-\frac{a_k}{d_k}\left(1-\lambda_k/\Gamma \right) \right),  
\label{eq:optimal_time}
\end{equation}
as shown in Refs.~\cite{busiello2021inducing,bao2022acceleratingrelaxationmarkovianopen}. The expression for the resetting time from \eqref{eq:optimal_time} must be physical, i.e., real and positive $t_r>0$. This requires $\lambda_k$ being real and additionally $a_k/d_k<0$. The latter thus provides a sufficient condition to observe strong Mpemba effect with unconditional resetting. Fullfillment of this condition, however requires fine tuning of both the initial and reset states and we will therefore not focus on this in order to induce, instead, relaxation speedup in a universal way.
In the main text, as well as in all the following sections of the Supplemental material, we quantify the relaxation dynamics by computing the distance between the state $\rho(t)$ of the system at time $t$ and the targeted stationary state $\rho_{\mathrm{ss}}$ of $\mathcal{L}_0$. The associated measure is denoted by $\bar{F}(\rho(t),\rho_{\mathrm{ss}})$ and it is given by: 
\begin{equation}
    \bar{F}(\rho(t),\rho_{\text{ss}})=1-F(\rho(t),\rho_{\text{ss}}).
\label{eq:bar_fidelity}
\end{equation}
Here, $F(\rho(t),\rho_{\text{ss}})$ is the fidelity between $\rho(t)$ and $\rho_{\mathrm{ss}}$, which is defined as
\begin{equation}
F(\rho(t),\rho_{\text{ss}})=\left(\text{Tr}\left[\sqrt{\sqrt{\rho(t)}\rho_{\text{ss}}\sqrt{\rho(t)}}\right]\right)^2.
\label{eq:fidelity}
\end{equation}
\section{Unconditional reset protocol for a qubit coupled to a thermal bath}
\label{sec:unconditional_qubit}
\begin{figure}
    \centering
    \includegraphics[width=0.75\linewidth]{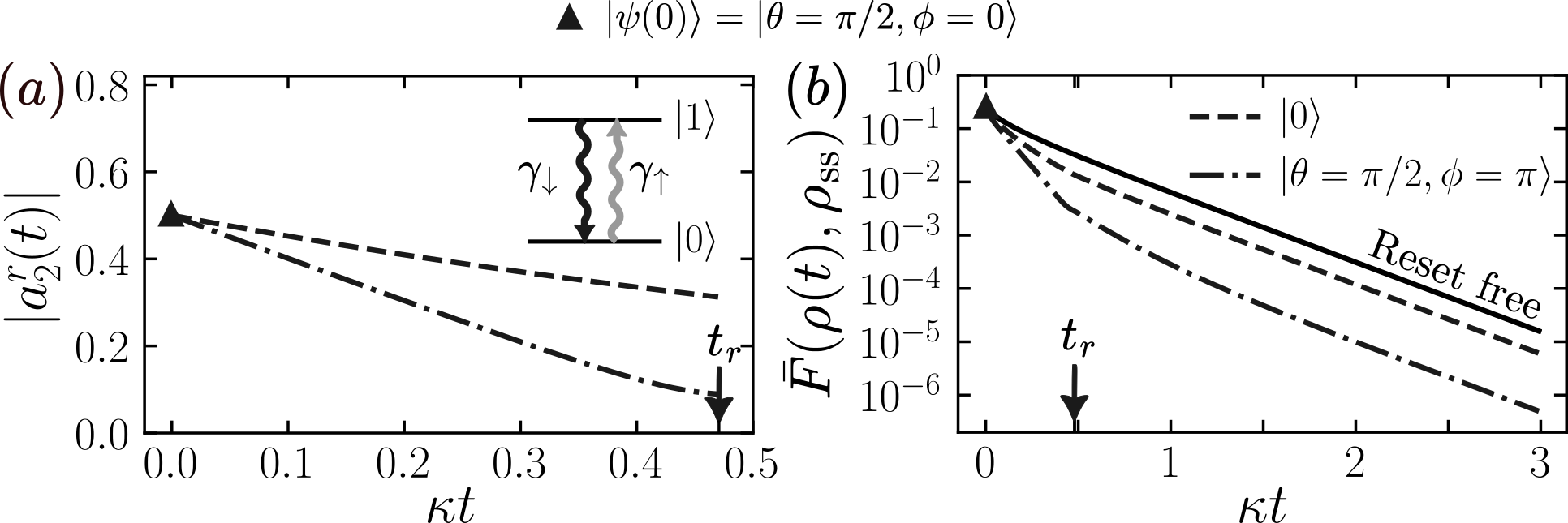}
    \caption{\textbf{Weak Mpemba effect in qubit.} 
     Initial state is reported at the top. Solid lines
indicate dynamics without resets, while dashed and dashed-dotted lines represent dynamics with resets to different states $\ket{\psi_r}$ (see labels) up to time $t_r$ (indicated by arrows). (a) Modified overlap $|a_2^r(t)|$ for a qubit coupled to a thermal bath (sketched in the inset). (b) Corresponding distance measure $\bar{F}(\rho(t),\rho_{\text{ss}})$, illustrating WME where $a_2^r(t)$ reaches a minimum under resetting.
    Parameters are $\omega/\kappa=1,n_{th}/\kappa=1$ where $\gamma_\downarrow=\kappa (1+n_{th})$ and $\gamma_\uparrow=\kappa n_{th}$. }
    \label{fig:figS1}
\end{figure}

We specialize the general method introduced in the previous section to the case of a qubit coupled to a thermal bath. 
This is described by the Hamiltonian $H=\omega \sigma_z$, where $\sigma_z=(\ket{1}\bra{1}-\ket{0}\bra{0})/2$ with $\ket{0},\ket{1}$ denoting the ground and excited state, respectively. Dissipation describes emission and absorption of photons to/from the bath with jump operators $\mathcal{O}_1=\sqrt{\gamma_\downarrow}\sigma_-$ and $\mathcal{O}_2=\sqrt{\gamma_\uparrow}\sigma_+$, respectively, as illustrated in the inset of Fig.~\ref{fig:figS1}(a). Here, $\sigma_-=\sigma_+^\dagger=\ket{0}\bra{1}$ and dissipation rates are fixed as $\gamma_\downarrow=\kappa (1+n_{th})$ and $\gamma_\uparrow=\kappa n_{th}$, where  $n_{th}=1/(e^{\beta \omega}-1)$ is given by the Bose-Einstein distribution at inverse temperature $\beta$. 
The associated  reset-free Liouvillian $\mathcal{L}_0$ is written as
\begin{equation}   \mathcal{L}_0(\rho)=-i[\omega \sigma_z, \rho]+\underbrace{\kappa (1+n_{th})}_{\gamma_{\downarrow}}D[\sigma_-]\rho+\underbrace{\kappa n_{th}}_{\gamma_{\uparrow}}D[\sigma_+]\rho.
\label{eq:reset_free_qubit}
\end{equation}

The reset-free dynamics therefore satisfies detailed balanced \cite{breuer2002} and it leads to thermalization of the qubit to a Gibbs state $\rho_{\mathrm{ss}}\sim \mbox{exp}(-\beta H)$.
The Liouvillian $\mathcal{L}_0$ is represented, after vectorialization (stacking rows) of the master equation \eqref{eq:reset_free_qubit}, as a $\mbox{dim}^2(\mathcal{H}) \times \mbox{dim}^2(\mathcal{H}) =4\times 4$ matrix:
\begin{equation}
\mathcal{L}_0=\begin{pmatrix}
    -\gamma_\downarrow & 0 & 0 & \gamma_\uparrow \\
    0 & -\frac{\gamma_\uparrow+\gamma_\downarrow}{2}+i \omega & 0 & 0\\
    0& 0 & -\frac{\gamma_\uparrow+\gamma_\downarrow}{2}-i \omega  & 0\\
     \gamma_\downarrow & 0 & 0  & -\gamma_\uparrow
\label{eq:qubit_matrix}
\end{pmatrix}.
\end{equation}
The $\mathcal{L}_0$ has the following four eigenvalues: 
\begin{equation}    \lambda_1=0,~~\lambda_{2}=\lambda_{3}^*=-\frac{\gamma_\uparrow+\gamma_\downarrow}{2}+ i \omega,~~\lambda_4=-(\gamma_\uparrow+\gamma_\downarrow).\label{eq:eigs_qubit}
\end{equation}
The slowest decaying modes $\lambda_2$ and $\lambda_3$ form in this case a complex conjugate pair. The right and left eigenvectors of the matrix \eqref{eq:qubit_matrix} are
\begin{align}
    R_1& =\frac{1}{\gamma_\uparrow+\gamma_\downarrow}\{\gamma_\uparrow,0,0,\gamma_\downarrow \}^T,~~ L_1=\{1,0,0,1\}^T, \nonumber\\
    R_2& =\{0,1,0,0\}^T, ~~L_2=\{0,0,1,0\}^T, \nonumber\\
    R_3& =\{0,0,1,0\}^T, ~~L_3=\{0,1,0,0\}^T, \nonumber\\
    R_4& =\{-1,0,0,1\}^T, ~~L_4=\frac{1}{\gamma_\uparrow+\gamma_\downarrow}\{-\gamma_\downarrow,0,0,\gamma_\uparrow\}^T.
\label{eq:eigenmodes_qubits}
\end{align}
By following the vectorization rules, the given eigenvectors can be expressed in matrix form as follows:
\begin{align}
    R_1& =\frac{1}{\gamma_\uparrow+\gamma_\downarrow}\begin{pmatrix}
    \gamma_\uparrow & 0 \\ 0 & \gamma_\downarrow \end{pmatrix},~~ L_1=\begin{pmatrix}
    1 & 0 \\ 0 & 1 \end{pmatrix}, \nonumber\\
    R_2& =\begin{pmatrix}
    0 & 1 \\ 0 & 0 \end{pmatrix}, ~~L_2=\begin{pmatrix}
    0 & 0 \\ 1 & 0 \end{pmatrix}, \nonumber\\
    R_3& =\begin{pmatrix}
    0 & 0 \\ 1 & 0 \end{pmatrix}, ~~L_3=\begin{pmatrix}
    0 & 1 \\ 0 & 0 \end{pmatrix}, \nonumber\\
    R_4& =\begin{pmatrix}
    -1 & 0 \\ 0 & 1 \end{pmatrix}, ~~L_4=\frac{1}{\gamma_\uparrow+\gamma_\downarrow}\begin{pmatrix}
    -\gamma_\downarrow & 0 \\ 0 & \gamma_\uparrow \end{pmatrix},
\label{eq:eigenmodes_qubits_2}
\end{align}
such that $\mbox{Tr}[L_i^\dagger R_j]=\delta_{ij}$ and $\mbox{Tr}[R_k]=\delta_{k1}$.

Resetting dynamics is implemented by adding the two jump operators 
\begin{equation}
\mathcal{O}^r_0=\sqrt{\Gamma}\ket{\psi_r}\bra{0}, \quad \mbox{and} \quad \mathcal{O}^r_1=\sqrt{\Gamma}\ket{\psi_r}\bra{1}.
\end{equation}

\textcolor{black}{In order to show concrete results, we begin by parametrizing the initial pure state on the Bloch sphere as $\ket{\psi(0)}=\ket{\theta,\phi}=\cos{(\theta/2)}\ket{1}+e^{i\phi}\sin{(\theta/2)}\}\ket{0}$, with $0 \leq \theta\leq \pi$ and $0 \leq \phi\leq 2\pi$. Any such initial state with $\theta \neq 0,\pi$ will result in $a_{2}\{=(a_{3})^*\} \neq 0$.
Thus, the relaxation to the steady state is governed by $Re[\lambda_{2,3}]=-(\gamma_\uparrow+\gamma_\downarrow)/2$, see Eq.~(\ref{eq:eigs_qubit}). }

\textcolor{black}{We note that for any reset state $\ket{\psi_r}$ that is diagonal in the qubit computational $z$ basis, one has  from the expression of $L_2,L_3$ in Eq.~\eqref{eq:eigenmodes_qubits} that $d_{2,3}=0$, and therefore from Eq.~\eqref{eq:overlaps} the modified overlap $a_{2,3}^r(t)$ is given by
\begin{equation}
a_{2,3}^{r}(t)=a_{2,3} e^{-\Gamma t}
\label{eq:overlap_qubit_diagonal}.
\end{equation}
The modified overlap $a_k^r(t_r)$ with the second and third eigenmodes of the Liouvillian is therefore reduced exponentially as a function of the time $t_r$ the resetting is applied. This is specifically the case of Fig.~\ref{fig:figS1}(a) and (b) for the reset state $\ket{\psi_r}=\ket{0}$. }

\textcolor{black}{For reset states that display coherences in the $z$ basis, on the contrary, $d_{2,3} \neq 0$. Interestingly, quantum coherence in such reset states can surprisingly enhance the relaxation process even for a classical process like thermalization of a qubit. This results from the fact that different terms in Eq.~(\ref{eq:akr}) with opposite sign can reduce $a_k^r(t_r)$ much faster than the diagonal reset state with $d_{2,3} = 0$. One such example is shown in Fig.~\ref{fig:figS1}(a) and (b), where a reset state $\ket{\psi_r}=\ket{\theta=\pi/2,\phi=0}$ results in faster convergence to steady state compared to the diagonal reset state $\ket{\psi_r}=\ket{0}$.}

\textcolor{black}{Unconditional reset to the ground state $\ket{0}$ can be experimentally realized by first performing a projective measurement in the $\{\ket{0},\ket{1}\}$ basis and subsequently applying $\pi$-pulse if the qubit is measured in the state $\ket{1}$. This idea can be extended to prepare the $\ket{x}=\ket{\theta=\pi/2,\phi=0}$ state 
by applying a suitable single-qubit gate, e.g., the Hadamard gate \cite{nielsen2010quantum}, after the measurement.}

\section{UNCONDITIONAL RESET PROTOCOL FOR A QUTRIT WITH METASTABLILITY}
\label{sec:unconditional_3_states}

\begin{figure}
    \centering
    \includegraphics[width=0.75\linewidth]{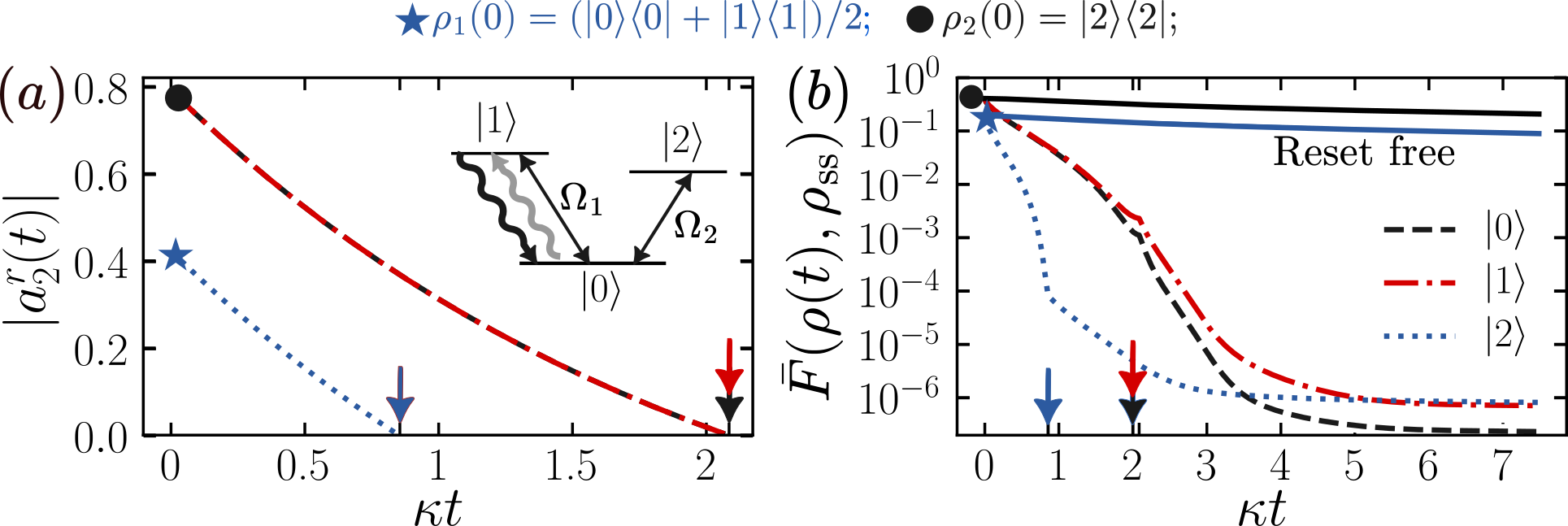}
    \caption{\textbf{Strong Mpemba effect in qubit.} 
     Initial states are reported at the top. Solid lines
indicate dynamics without resets, while dashed and dashed-dotted lines represent dynamics with resets to different states $\ket{\psi_r}$ (see labels) up to time $t_r$ (indicated by arrows). (a) Modified overlap $|a_2^r(t)|$ for the qutrit (see inset). (b) Distance measure $\bar{F}(\rho(t),\rho_{\text{ss}})$, depicting SME where $a_2^r(t_r)=0$ for multiple reset states, yielding exponential speedup toward the steady state. Parameters are $\Omega_1/\kappa=1,\Omega_2/\kappa=0.1,n_{th}/\kappa=2$, where $\gamma_\downarrow=\kappa (1+n_{th})$ and $\gamma_\uparrow=\kappa n_{th}$. }
    \label{fig:figS2}
\end{figure}

\textcolor{black}{We extend our analysis of unconditional resetting to a three-level system, illustrated in the inset of Fig.~\ref{fig:figS2}(a).
This system features two coherent laser-drives between the states  $0 \leftrightarrow 1 $, at frequency $\Omega_1$, and  $0 \leftrightarrow 2 $, at frequency $\Omega_2$ and the 
Hamiltonian is $H=\Omega_1 \ket{0}\bra{1}+\Omega_2 \ket{0}\bra{2}+\mathrm{h.c.}$. 
The levels $\ket{0}$ and $\ket{1}$ are still coupled to a thermal bath at inverse temperature $\beta$ and consequently subject to emissions/absorptions of photons in the bath, with jump operators $\mathcal{O}_1=\sqrt{\gamma_\downarrow} \ket{0}\bra{1}$ and $\mathcal{O}_2=\sqrt{\gamma_\uparrow}\ket{1}\bra{0}$ and the rates $\gamma_\downarrow=\kappa (1+n_{th})$ and $\gamma_\uparrow=\kappa n_{th}$. 
Due to the additional level $\ket{2}$ and the laser driving, detailed balance is nevertheless broken for this model. 
Relaxation takes place towards a mixed non-equilibrium stationary state with a finite population on each level. 
Timescales for reaching stationarity are, however, extremely slow for $\Omega_1\gg \Omega_2$ when the state $\ket{2}$ is metastable, known as the ``shelving" configuration \cite{plenio1998quantumjumps,zoller1987,open_meta1,open_meta2}. 
This reflects into a very small spectral gap determined by $\lambda_2$.}

\textcolor{black}{Transient resetting has a dramatic effect here, leading to SME and an exponential speedup of relaxation (see Fig.~\ref{fig:figS2}(a) and (b)). 
To showcase this effect, we consider two different initial states with finite $a_2$. One belongs to the fast subspace spanned by $\ket{0}$ and $\ket{1}$:  $\rho_1(0)=(\ket{0}\bra{0}+\ket{1}\bra{1})/2$; the other one lies in the complementary slow subspace given by $\rho_2(0)=\ket{2}\bra{2}$. 
As in the qubit case, the choice of the reset state is solely dictated by the finite inverse temperature $\beta$ of the $0 \leftrightarrow 1$ line. We then consider resetting to the complementary subspaces for each initial state, i.e, we reset to the state $\ket{\psi_r}=\ket{2}$ for initial state  $\rho_1(0)$ and to $\ket{\psi_r}=\ket{0}$ or $\ket{1}$ for $\rho_2(0)$. An SME effect can be induced for both these initial states such that $|a_2^r(t_r)|=0$, cf. Fig.~\ref{fig:figS2}(a). As a result, metastable slow-dynamics is filtered out after $t_r$, and the system converges exponentially faster to the steady state (see Fig.~\ref{fig:figS2}(b)). }

\begin{figure}[htp!]
    \centering
\includegraphics[width=0.8\linewidth]{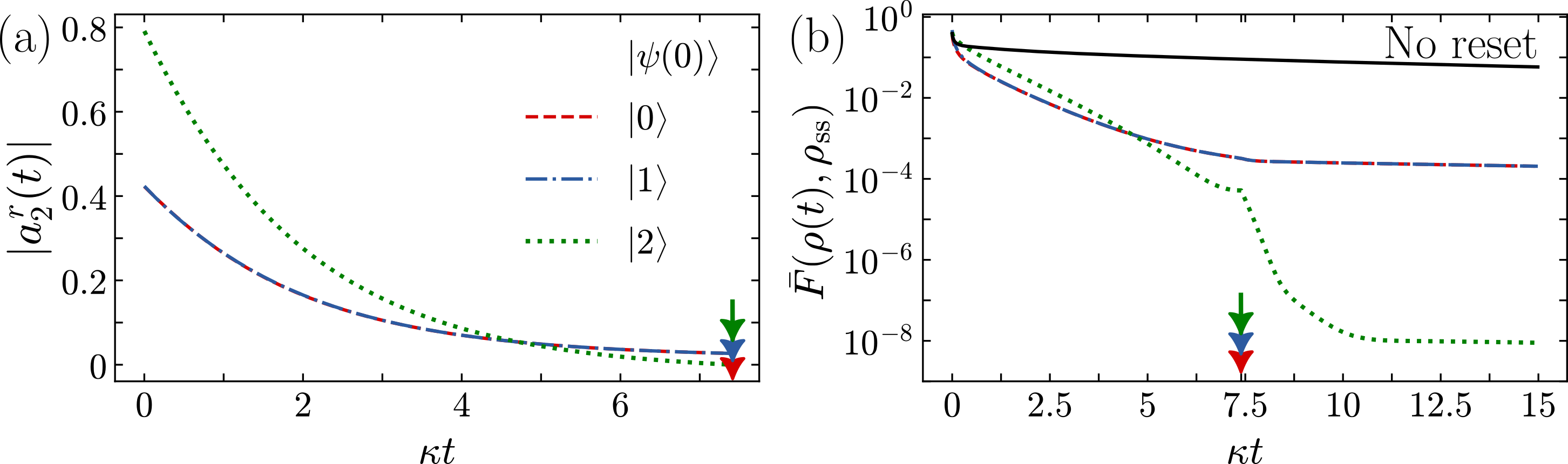}
    \caption{Mpemba effect in a qutrit system with unconditional resetting to a mixed state infinite-temperature state $\mu_r=\mathbb{I}/3$. In panel (a) we plot the coefficient $|a_2^r(t_r)|$ Eq.~\eqref{eq:akr} as a function of the rescaled time $\kappa t$, with $\kappa$ the dissipation rate of $\ket{0}-\ket{1}$ line. We take various initial states $\ket{\psi(0)}$ (dashed lines). For $\ket{\psi(0)}=\ket{2}$, the overlap $|a_2^r(t_r)|=0$ and strong Mpemba effect is found. For the other initial states $\ket{0}$ and $\ket{1}$, the overlap $|a_2^r(t_r)|$ is minimized, but it remains nonzero. This reflects the occurrence of the weak Mpemba effect. In panel (b), we plot the distance measure $\bar{F}(\rho(t),\rho_{\text{ss}})$ Eq.~\eqref{eq:bar_fidelity} and Eq.~\eqref{eq:fidelity} for initial states considered in (a). For all the initial states, relaxation is significantly accelerated compared to the reset-free dynamics (black solid line), which shows metastable, very slow convergence to stationarity. The parameter values are as follows: $\Omega_1/\kappa=1,\Omega_2/\kappa=0.1,n_{th}/\kappa=2$ and $\Gamma=1/2$. Arrows indicate the time $t_r$ when stochastic resets are turned off.}
    \label{fig:mixed_state}
\end{figure}

\textcolor{black}{This effect is not specific to this choice of initial states. We, instead, show that SME is also found for states easy to experimentally access, such as mixed infinite-temperature states. }
We denote it as $\mu_r$. The dynamics of the system under such resetting is obtained by simply generalizing Eq.~\eqref{eq:reset} to the case of a mixed reset state
\begin{equation}   \dot\rho_r=\mathcal{L}_r(\rho_r)=\mathcal{L}_0(\rho_r)+\Gamma \, \mbox{Tr}(\rho_r) \mu_r-\Gamma  \rho_r.
\end{equation}
For the qutrit system described above, the mixed reset state is a statistical mixture of the three energy eigenstates $\ket{0}$, $\ket{1}$, and $\ket{2}$ as
\begin{equation}
\mu_r = \sum_{j=0}^{2} p_j \ket{j}\bra{j}, 
\label{eq:mixed_resetting}
\end{equation}
where the probabilities $p_j$ satisfy the normalization $\sum_{0=1}^{2} p_j = 1$.
Resetting to a mixed state $\mu_r$ can be obtained via considering the following set of nine jump operators $\mathcal{O}^r_{j,k}=\sqrt{\Gamma p_j}\ket{j}\bra{k}$ where $j,k\in \{0,1,2\}$. In Fig.~\ref{fig:mixed_state}, we report the results for unconditional resetting to a mixed state Eq.~\eqref{eq:mixed_resetting} with $p_j=p=1/3\,\, \forall j$. We choose this state since it can be readily experimentally realized, e.g., by coupling the system to an infinite-temperature bath. 

In Fig.~\ref{fig:mixed_state}(a), we plot the modulus $|a_2^r|$ of the modified overlap Eq.~\eqref{eq:akr} as a function of time for various initial states $\ket{\psi(0)}$. If the system is initialized in the states $\ket{0}$ and $\ket{1}$, a weak Mpemba effect is observed since the modified overlap can be minimized, but not canceled $\vert a_2^r(t_r) \vert \neq 0$. This results in faster relaxation compared to the no reset case, as shown in Fig.~\ref{fig:mixed_state}(b). For the initial state $\ket{2}$, instead, the system experiences a strong Mpemba effect since the resetting time $t_r$ can be fixed such that the overlap $|a_2^r(t_r)|=0$ is exactly canceled. This is shown in Fig.~\ref{fig:mixed_state}(a). One therefore finds exponential speed-up of relaxation, as displayed in Fig.~\ref{fig:mixed_state}(b). This analysis demonstrates that the occurrence of the Mpemba effect in a qutrit is not limited to pure reset states, which are discussed in the main text. The Mpemba effect, however, also occurs in cases of naturally occurring and easily accessible mixed reset states, such as infinite-temperature states. In this case, acceleration of relaxation is obtained for any initial state.

\section{Conditional resetting in qubit and qutrit \label{sec:conditional}}
\begin{figure}[t!]
    \centering   \includegraphics[width=0.8\linewidth]{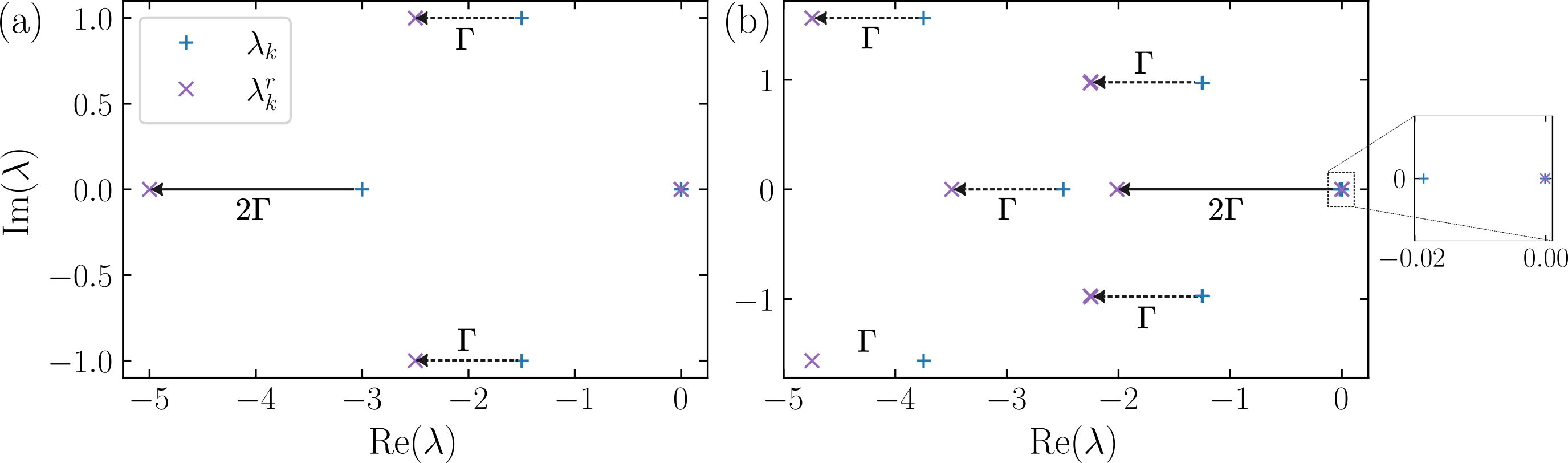}
    \caption{Eigenspectra of Liouvillians $\mathcal{L}$ for the qubit in panel (a) and qutrit in panel (b). The effect of conditional resetting leads to an uneven shift in eigenvalues for the conditional reset protocol compared to the eigenspectra of the reset-free Liouvillian. Eigenvalues of the conditional resetting maps are denoted with crosses `x', while eigenvalues of the reset-free corresponding generators with plus symbols `+'.}
    \label{fig:eigs}
\end{figure}

\begin{figure}[htp!]
    \centering
    \includegraphics[width=0.75\linewidth]{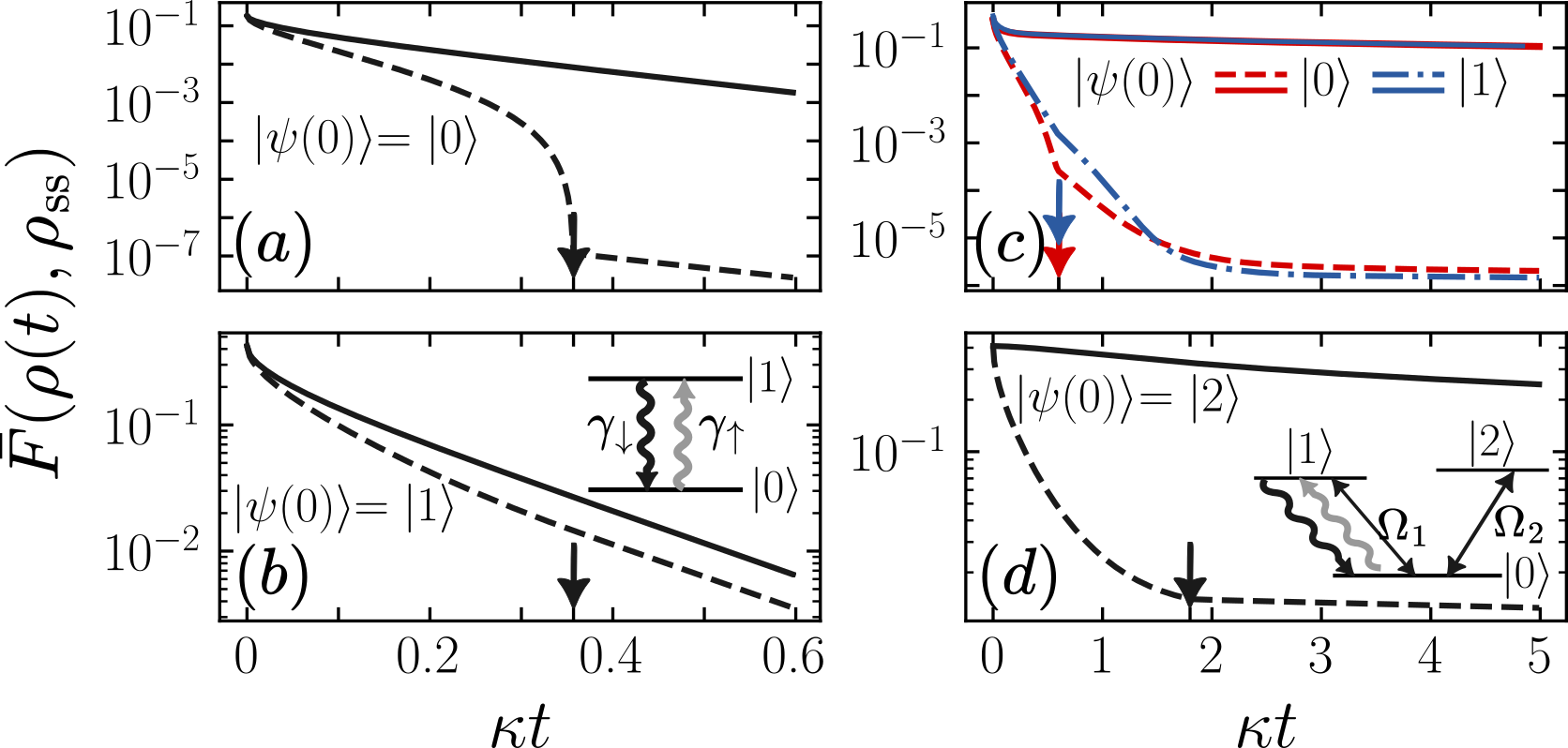}
    \caption{\textbf{Mpemba effect via conditional resetting.} In panels (a) and (b), we show the emergence of SME and WME, respectively, in a qubit. Panels (c) and (d) show the SME and WME in a qutrit, respectively. Different initial states are reported, each labeled, while arrows mark the time $t_r$ up to which resets are performed. Black-solid lines correspond to the reset-free dynamics and are reported for comparison. Parameters for qubit (a,b) are $\omega/\kappa=1,n_{th}/\kappa=1$, and for qutrit (c,d) are $\Omega_1/\kappa=1,\Omega_2/\kappa=0.1,n_{th}/\kappa=2$, where $\gamma_\downarrow=\kappa (1+n_{th})$ and $\gamma_\uparrow=\kappa n_{th}$. }
    \label{fig:fig4}
\end{figure}

\textcolor{black}{In this section, we analyze conditional resetting in finite-level quantum systems. We start with a qubit coupled to a thermal bath, as discussed in previous sections. Subsequently, we examine a form of conditional resetting that functions as a thermostat. This protocol cools down the qubit to the ground state $\ket{0}$ if the measured energy is higher than that of the finite-temperature stationary state, and, vice versa, heats up the system to $\ket{1}$ if the measured energy is lower. There are, accordingly, two reset states $\mathcal{R}=2$ in Eq.~(4) of the main text: $\ket{\psi_0}=\ket{0}$ and $\ket{\psi_1}=\ket{1}$, which are chosen dependently on the overlap with the excited $P_0=\ket{1}\bra{1}$ and ground $P_1=\ket{0}\bra{0}$ sector of the Hilbert space, respectively. This conditional resetting operation can be implemented via the application of a bit-flip gate $\mathcal{R}_{\theta}^x$ \cite{rqh3,rqh4} (rotation of an angle $\pi$ around the $x$ axis), as sketched in Fig.~1(c) of main text. Equation (4) of the main text can be written in the Lindblad form by identifying the two jump operators }
\begin{equation}
\mathcal{O}^c_0=\sqrt{\Gamma}\vert 0 \rangle \langle 1 \vert, \quad \mbox{and} \quad \mathcal{O}^c_1=\sqrt{\Gamma}\vert 1 \rangle \langle 0 \vert, 
\label{eq:qubit_conditional_jumps}
\end{equation}
where $\mathcal{O}^c_\alpha$ models the conditional reset to the state $\ket{\alpha}$.
The conditional reset Liouvillian for the above case (Eq.~(4) of the main text) is consequently given by
\begin{equation}
\dot{\rho}=\mathcal{L}_r^c(\rho)=\mathcal{L}_0(\rho)+\Gamma (\bra{0}\rho \ket{0}\ket{1}\bra{1}+\bra{1}\rho \ket{1}\ket{0}\bra{0})-\Gamma \rho,
\label{eq:conditional_qubit_liovillian}
\end{equation}
where the reset-free Liouvillian $\mathcal{L}_0$ of the qubit is given by \eqref{eq:reset_free_qubit}. In the presence of the conditional resetting, the reset rates add to the thermal rates $\gamma_{\uparrow,\downarrow} \rightarrow \gamma_{\uparrow,\downarrow}+\Gamma$, such that the Liouvillian in vectorized representation has the form
\begin{equation}
\mathcal{L}_r^c=\begin{pmatrix}
    -\gamma_\downarrow-\Gamma & 0 & 0 & \gamma_\uparrow+\Gamma \\
    0 & -\frac{\gamma_\uparrow+\gamma_\downarrow}{2}-\Gamma+i \omega_0 & 0 & 0\\
    0 & 0 & -\frac{\gamma_\uparrow+\gamma_\downarrow}{2}-\Gamma-i \omega_0  & 0\\
     \gamma_\downarrow+\Gamma & 0 & 0  & -\gamma_\uparrow-\Gamma\\
\end{pmatrix}.
\end{equation}
In contrast to the unconditional resetting discussed in Sec.~\ref{sec:unconditional}, the eigenvalues of $\mathcal{L}_r^c$ do not follow from a constant shift of $-\Gamma$ from the eigenvalues of the reset-free Liouvillian \eqref{eq:reset_free_qubit} and \eqref{eq:qubit_matrix}; rather, different eigenvalues experience different shifts.
The eigenvalues in the presence of resetting are, indeed, given by: 
\begin{equation}   \lambda_1^r=0,~~\lambda_{2,3}^r= -\Gamma+\lambda_{2,3},~~\lambda_4^r=-2\Gamma+\lambda_4.
\label{eq:conditional_nonconstant_shift}
\end{equation}
This shift in eigenvalues is depicted in Fig.~\ref{fig:eigs}(a). Similar to the unconditional reset, the right eigenvectors do not change under conditional resetting except $R_1$ (steady-state).
This can be verified by explicitly checking
\begin{equation}
\mathcal{L}_r^c(R_k)=\lambda^r_k R_k.
\label{eq:conditional_R_eigenmatrix}
\end{equation}
The steady state $R_1^r$ in the presence of resetting gets modified 
\begin{align}
    R_1^r =\frac{1}{\gamma_\uparrow+\gamma_\downarrow+2\Gamma}\{\gamma_\uparrow+\Gamma,0,0,\gamma_\downarrow+\Gamma \}^T,
\end{align}
as $\gamma_{\uparrow,\downarrow} \rightarrow \gamma_{\uparrow,\downarrow}+\Gamma$.
Accordingly, we can check that the left eigenvectors are modified as 
\begin{equation}
    L_k^r=L_k+\alpha_k I
\end{equation}
where $\alpha_k$ can be found by using $(\mathcal{L}_r^c)^\dagger(L_k^r)=(\lambda^r_k)^* L_k^r$.
In particular, $\alpha_{2,3}=0$ and $\alpha_4=\frac{\Gamma (\gamma_\uparrow-\gamma_\downarrow)}{(\gamma_\uparrow+\gamma_\downarrow)(\gamma_\uparrow+\gamma_\downarrow+2\Gamma)}$. From the orthogonality relation $Tr[(L^r_k)^\dagger R_1^r]=\delta_{k1}$ and the expansion of the steady state in the right reset-free eigenbasis, one eventually has
\begin{equation}
    R_1^r=R_1+\sum_{k\neq 1}c_kR_k=R_1^r=R_1-\alpha_4 R_4,
\end{equation}
such that 
\begin{equation}
c_{2,3}=0, \quad \mbox{and} \quad c_4=-\alpha_4=-\frac{\Gamma (\gamma_\uparrow-\gamma_\downarrow)}{(\gamma_\uparrow+\gamma_\downarrow)(\gamma_\uparrow+\gamma_\downarrow+2\Gamma)}.
\label{eq:conditional_c_coefficient}
\end{equation}
With the information in Eqs.~\eqref{eq:conditional_R_eigenmatrix}-\eqref{eq:conditional_c_coefficient} at our disposal, we can eventually write the time evolution of the density matrix $\rho^r(t)$ in the presence of conditional resetting in terms of the reset-free overlaps and eigenvalues:
\begin{align}
\rho^r(t)&=R^r_1+\sum_{k\neq 1}e^{\lambda_k^rt}\mbox{Tr}[(L_k^r)^\dagger \rho(0)]R_k^r \nonumber \\
    &=R_1+\sum_{k\neq 1}c_kR_k+\sum_{k\neq 1}e^{\lambda_k^rt}\mbox{Tr}[(L_k^r)^\dagger \rho(0)]R_k \nonumber \\
    &=R_1+\sum_{k\neq 1}\left(c_k e^{-\lambda_k t} +e^{(\lambda_k^r-\lambda_k)t}\mbox{Tr}[(L_k^r)^\dagger \rho(0)]\right)e^{\lambda_k t}R_k \nonumber \\
    &=R_1+\sum_{k\neq 1}\left(c_k e^{-\lambda_k t} +e^{(\lambda_k^r-\lambda_k)t}\{\mbox{Tr}[(L_k)^\dagger \rho(0)]+\alpha_k\}\right)e^{\lambda_k t}R_k \nonumber \\
    &=R_1+\sum_{k\neq 1}a^r_k(t)e^{\lambda_k t}R_k,
\end{align}
where we defined the modified overlap $a_k^r(t)$ due to conditional resetting as 
\begin{align}
    a_k^r(t)&=c_k e^{-\lambda_k t}+e^{(\lambda_k^r-\lambda_k)t}\{\mbox{Tr}[(L_k)^\dagger \rho(0)]-c_k\}=c_k e^{-\lambda_k t}+e^{(\lambda_k^r-\lambda_k)t}(a_k-c_k)\nonumber \\
    &=c_k(e^{-\lambda_k t}-e^{(\lambda_k^r-\lambda_k) t})+a_k e^{(\lambda_k^r-\lambda_k) t}.
\label{eq:conditional_overlap}
\end{align}
The relation between the eigenvalues $\lambda_k^r$ of the resetting map and those $\lambda_k$ of the reset-free map is reported in Eq.~\eqref{eq:conditional_nonconstant_shift}. The definition of the reset-free overlaps $a_k$ is reported in Eq.~\eqref{eq:overlaps}. Similar to the case of unconditional resetting, we can now use tune the resetting time $t_r$ to either induce SME by exactly canceling the overlap $a^r_k(t_r)=0$, or the WME by minimizing it $\vert a^r_k (t_r)\vert$. Strong Mpemba effect is readily obtained by finding the time $t_r$ that the right hand side of \eqref{eq:conditional_overlap} is zero. This leads to 
\begin{equation}
t_r=-\frac{1}{\lambda_k^r}\ln \left(1-\frac{a_k}{c_k}  \right),   
\label{eq:conditional_qubit_optimal_time}
\end{equation}
which provides a physical solution for $t_r$ if 
\begin{equation}
a_k/c_k<1,
\label{eq:sufficient_conditional}
\end{equation}
and $\lambda_k^r$ is real. Equation \eqref{eq:sufficient_conditional} is thus a sufficient condition for SME via conditional resetting in the qubit model. \textcolor{black}{In Fig.~\ref{fig:fig4}(a), we show the dynamics from the initial state $\ket{\psi(0)}=\ket{0}$, while in Fig.~\ref{fig:fig4}(b), the initial state $\ket{\psi(0)}=\ket{1}$ is taken. Speedup of relaxation is found for all the initial state. In particular a SME is observed for initial state $\ket{\psi(0)}=\ket{0}$ as shown in Fig.~\ref{fig:fig4}(a), while a WME is observed for initial state as depicted in \ref{fig:fig4}(b).}

In the case of a qutrit in the conditional resetting protocol there are $\mathcal{R}=3$ three possible reset states: $\ket{0}$, $\ket{1}$ and $\ket{2}$. The reset state is chosen again based on a measurement of the energy. If the qutrit is energy is that of the metastable state $\ket{2}$, then the system is reset to $\ket{0}$ or $\ket{1}$ with the same probability. Otherwise, if the energy is measured to be that of the state $\ket{0}$ or $\ket{1}$, then the qutrit is reset to the metastable state $\ket{2}$. The Liouvillian in Eq.~(4) of the main text is then described by the following operators $P_r$, with $r=1,2,3$:
\begin{equation}
P_0=\frac{1}{2}\ket{2}\bra{2}   \quad P_1= \frac{1}{2}\ket{2}\bra{2}, \quad P_2=\ket{0}\bra{0}+\ket{1}\bra{1}.
\label{eq:projectors_qutrit}
\end{equation}
We note that $P_0$ and $P_1$ are not projectors ($P_{0,1}^2 \neq P_{0,1}$). This simply follows from our choice that when the qutrit is measured in the metastable state $\ket{2}$, it is reset to $\ket{0}$ or $\ket{1}$ with the same probabilities $1/2$. The fundamental property $\sum_{r=0}^2 P_r=\mathbb{I}$ holds, which represents completeness of the energy measurement. The Liouvillian in Eq.~(4) of the main text can be written in the Lindblad form by defining from Eq.~\eqref{eq:projectors_qutrit} the following four jump operators
\begin{equation}
\mathcal{O}^c_{02}=\sqrt{\Gamma/2}\vert 0 \rangle \langle 2 \vert, \quad \mathcal{O}^c_{12}=\sqrt{\Gamma/2}\vert 1 \rangle \langle 2 \vert, \quad\mathcal{O}^c_{20}=\sqrt{\Gamma}\vert 2 \rangle \langle 0 \vert, \quad \mbox{and} \quad  \mathcal{O}^c_{21}=\sqrt{\Gamma}\vert 2 \rangle \langle 1 \vert.
\label{eq:qutrit_conditional_jumps}
\end{equation}
In the qutrit case, we find that both the left and right eigenvectors change, which means that we cannot obtain an analytical form for the overlap $a_k^r(t)$ comparable to \eqref{eq:conditional_overlap} for the qubit case. We, however, numerically observe that the eigenvalues undergo an uneven shift, similar to the behavior exhibited by qubits. This is reported in Fig.~\ref{fig:eigs}(b). Notably, in this protocol, the metastable eigenvalue $\lambda_1$ is shifted by $2\Gamma$, while the other eigenvalues undergo a constant shift of $\Gamma$. We can still numerically calculate the overlap $a_k^r$ by expanding the state $\rho^r(t)$ in the presence of resetting into the reset-free eigenbasis of $\mathcal{L}_0$ as
\begin{equation}
a_k^r(t)=\mbox{exp}(-\lambda_k t) \mbox{Tr}[L_k^\dagger \rho^r(t)]. 
\end{equation}
This expression is exploited to check whether WME and SME can be induced with such conditional reset protocols in Fig.~\ref{fig:fig4}(c,d). \textcolor{black}{This conditional resetting protocol leads to a Mpemba effect for all the initial states.
In particular, the system exhibits a SME if it is initialized in the $\ket{0}$ and $\ket{1}$ subspaces as shown in Fig.~\ref{fig:fig4}(c), while the WME is observed when the initial state is $\rho(0) = \ket{2}\bra{2}$; see  Fig.~\ref{fig:fig4}(d).}

\section{Conditional vs unconditional reset \label{sec:cond_vs_uncond}}
In this section, we compare the conditional and unconditional reset protocols for a qubit coupled to a thermal bath. To ensure a fair comparison between the two protocols, we take the same reset rate $\Gamma$ for both. This ensures from Eqs.~\eqref{eq:qubit_conditional_jumps} and \eqref{eq:qutrit_conditional_jumps} that $\sum_i (\mathcal{O}_i^c)^\dagger \mathcal{O}_i^c = \Gamma \mathbb{I} $, where $\mathbb{I}$ is the identity matrix. The escape rate, i.e., the rate of performing a reset (regardless of the chosen reset state), is therefore the same as in the unconditional resetting, where
$\mathcal{O}_i^r=\sqrt{\Gamma}\ket{\psi_r}\bra{\phi_i}$ and $\sum_i (\mathcal{O}_i^r)^\dagger \mathcal{O}_i^r = \Gamma \mathbb{I} $. This ensures that for the same resetting time $t_r$, the same number of reset events take place, on average, in the two protocols.

\begin{figure}[h!]
    \centering
    \includegraphics[width=1\linewidth]{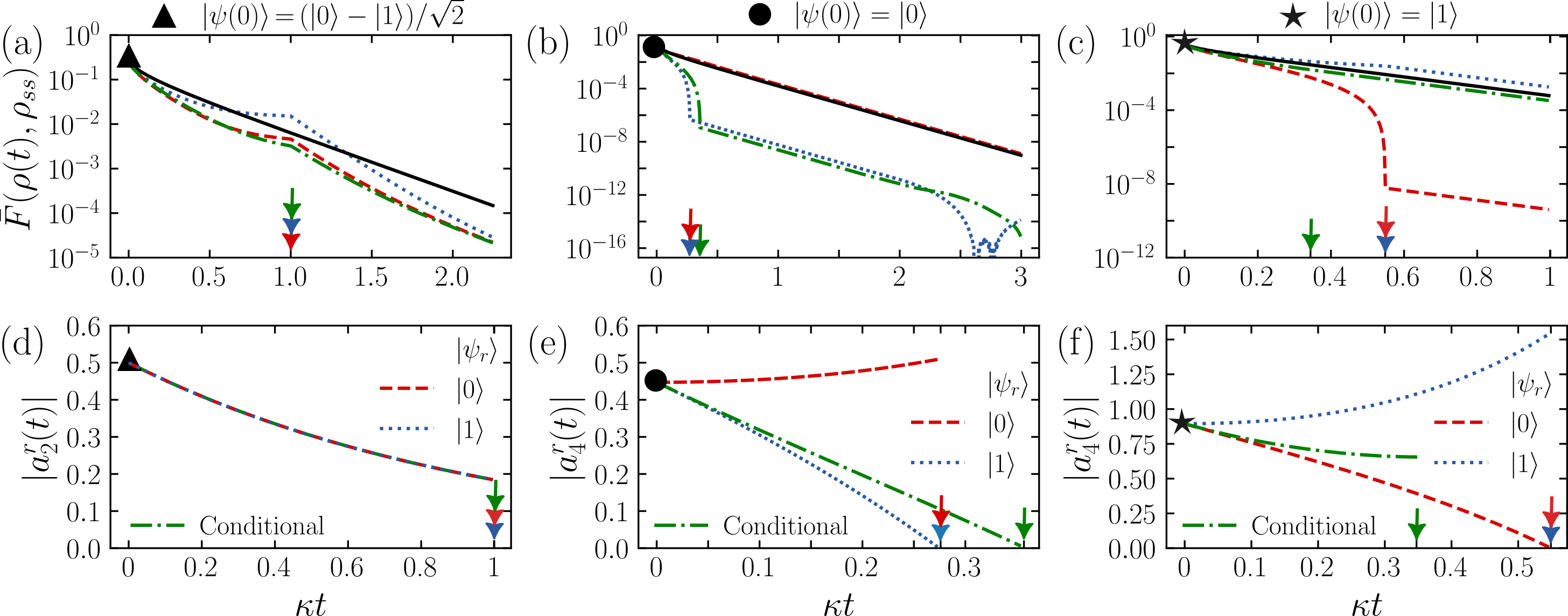}
    \caption{Panels (a), (b), and (c) show the distance measure $\bar{F}(\rho(t), \rho_{\text{ss}})$ in Eqs.~\eqref{eq:bar_fidelity} and \eqref{eq:fidelity} as a function of the rescaled time $kt$. This is calculated for various initial states $\rho = \ket{\psi_0}\bra{\psi_0}$, which are reported on top of the panels. Reset states for the unconditional resetting protocol are reported in the legend and they correspond to the dashed curves. Conditional resetting is represented by the dashed-dotted red line. For both protocols, arrows indicate the time $t_r$ up to which resetting is applied. The solid line represents the time evolution of the reset-free state for all initial states. Panels (d), (e), and (f) illustrate the time evolution of the overlap $a_k^r(t)$ for different reset protocols, both conditional and unconditional, associated with panels (a), (b), and (c), respectively. Note that in panel (d), the slowest decaying mode is $k=2$ since $|a_2^r(t)| \neq 0$. In panels (e) and (f), instead, the slowest decaying mode is $k=4$ since $a_2^r(t)=a_3^r(t)=0$ for both unconditional and conditional resetting. The parameters used are: $\omega_0/\kappa = n_{th}/\kappa =\Gamma/\kappa=1$.}
    \label{fig:cond_vs_uncond}
\end{figure}

We begin in Fig.~\ref{fig:cond_vs_uncond}(a) and (d) by considering an initial state with non-vanishing coherences in the energy eigenbasis, such that $a_{2,3}\neq 0$ in Eq.~\eqref{eq:overlaps} (with $L_k$ in Eq.~\eqref{eq:eigenmodes_qubits}). Specifically, we take $\ket{\psi(0)}=(\ket{0}-\ket{1})/\sqrt{2}$. For the unconditional reset protocol, we examine two different reset states $\ket{\psi_r}$: $\ket{0}$ and $\ket{1}$. The conditional protocol dynamics is dictated by Eqs.~\eqref{eq:qubit_conditional_jumps} and \eqref{eq:conditional_qubit_liovillian}. In this case, we find that for both protocols $a_{2,3}^r(t) =a_{2,3}\mbox{exp}(-\Gamma t)$ since $d_{2,3}=0$ in \eqref{eq:overlaps} and $c_{2,3}=0$ in \eqref{eq:conditional_c_coefficient}. The two resetting protocols therefore identically modify the overlap $a_{2}^r(t)$ with the slowest decaying mode $k=2$, as shown in Fig.~\ref{fig:cond_vs_uncond}(d). The system displays WME since $a_2^r(t_r)\neq 0$. In Fig.~\ref{fig:cond_vs_uncond}(a), the fidelity $\bar{F}(\rho(t),\rho_{\mathrm{ss}})$ between $\rho(t)$ and the stationary state $\rho_{\mathrm{ss}}$ is, however, slightly smaller for conditional resetting, since in this case the overlap with the next subleading mode $a_4^r(t)$ is smaller compared to unconditional resetting.

Next, we analyze in Figs.~\ref{fig:cond_vs_uncond}(b) and (c) diagonal initial states in the energy eigenbasis, $\ket{\psi(0)}=\ket{0}$  and $\ket{\psi(0)}=\ket{1}$. In this case $a_{2,3}^r = 0$ for unconditional resetting since $a_{2,3}=0$ and $d_{2,3}=0$ according to Eq.~\eqref{eq:overlaps} and \eqref{eq:eigenmodes_qubits}. For conditional resetting, similarly, $a_{2,3}=0$ and $c_{2,3}=0$ in Eq.~\eqref{eq:conditional_c_coefficient}. Therefore $a_2^r=0$ in Eq.~\eqref{eq:conditional_overlap}. In both the protocols, we therefore study how resetting can suppress the overlap with the fourth excited mode $a_4^r$. This is shown in Figs.~\ref{fig:cond_vs_uncond}(e) and (f). Since the stationary state has a finite inverse temperature $\beta$, both the states $\ket{0}$ and $\ket{1}$ have to develop in time a finite population. Unconditional resetting is therefore capable of inducing SME only when the reset state is complementary to the initial state. For instance, an initial ground state $\ket{0}$ exhibits SME when reset state is $\ket{1}$, see Fig.~\ref{fig:cond_vs_uncond}(b) and (e), and vice versa, see Fig.~\ref{fig:cond_vs_uncond}(c) and (f). The unconditional reset protocol fails to generate any Mpemba effect and, conversely, slows down relaxation if the reset state coincides with the initial state.
\textcolor{black}{By contrast, the sufficient condition Eq.~(\ref{eq:sufficient_conditional}) is satisfied for $\ket{\psi(0)}=\ket{0}$ and the conditional reset protocol thus induces SME with $t_r$ as in Eq.~\eqref{eq:conditional_qubit_optimal_time}  (see Fig.~\ref{fig:cond_vs_uncond}(b) and (e)). For the state $\ket{\psi(0)}=\ket{1}$, we, instead, observe WME from conditional resetting (see Fig.~\ref{fig:cond_vs_uncond}(c) (f)).}

In conclusion, the effectiveness of the unconditional reset protocol depends on the choice of initial and reset states. These do not require fine-tuning and are fixed solely on the basis of the inverse temperature $\beta$ of the stationary state. Conditional resetting, on the other hand, exhibits more robust behavior across different initial conditions, as it accelerates relaxation for any chosen initial state. The latter only affects the nature of the Mpemba effect, whether it is strong or weak.

\section{Mpemba effect in Dicke model via unconditional resetting \label{sec:dicke}}
\label{sec:unconditional_Dicke}
We now consider the open Dicke model \cite{carollo2021mpemba,PhysRevA.99.033845,imai2019dynamical} as a paradigmatic example of an out-of-equilibrium many-body system describing interaction between light and atoms in a cavity. This model is known for exhibiting rich behavior, including transitions between superradiant and subradiant phases, as well as non-stationary states such as continuous-time crystals. The dynamics of the open Dicke model can be described by the following master equation
\begin{align}
    \dot{\rho}=-i[\omega \hat{S}_z+\omega_ca^{\dagger} a+\frac{g}{\sqrt{N}}(a+a^{\dagger})\hat{S}_x,\rho ]+\kappa D[a]\rho,
\end{align}
where $a$ is the annihilation operator for the cavity modes, $\hat{S}_\alpha=\sum_{k=1}^N\sigma_\alpha^k$ are the collective spin operators, with $\sigma_\alpha^i$ being the Pauli spin-operators and $\alpha\in \{x,y,z\}$. The model contains the four parameters: $\omega$, the atomic energy splitting, $\omega_c$, the photon energy, $g$, the light-atoms interaction constant, and $\kappa$ the loss rate of photons. In the limit where  $\kappa \gg \omega,\omega_c,g$,  we can adiabatically eliminate the cavity and obtain the spin-only description of the open Dicke model \cite{PhysRevA.99.033845,imai2019dynamical}, as follows:
\begin{equation}
    \dot{\rho}=-i[\omega \hat{S}_z -\frac{4 g^2 \omega_c}{N(4\omega_c^2+\kappa^2)}\hat{S}^2_x,\rho ]+\frac{4g^2\kappa}{N(4\omega_c^2+\kappa^2)} D[\hat{S}_x]\rho. \label{eq:dicke}
\end{equation}
Note that this effective master equation does not exhibit any phase transition in the steady state \cite{PhysRevA.99.033845,imai2019dynamical}.
This is because the steady state of Eq.~(\ref{eq:dicke}) is always fully mixed, $\rho_{\text{ss}}=\mathbb{I}/Tr[\mathbb{I}]$, independent of the parameters $g$, $\omega$, $\omega_c$ and $\kappa$. The total spin $\hat{S}^2=\hat{S}_x^2+\hat{S}_y^2+\hat{S}_z^2$ is conserved under the dynamics in Eq.~\eqref{eq:dicke} and we can therefore restrict the dynamics to the Dicke manifold $\ket{S,S_i}$ labeled by the quantum number $S$ associated to $\hat{S}^2$ and the quantum number $S_i$ associated to $\hat{S_z}$. We consider the sector of the Hilbert space henceforth with maximal total angular momentum $S=N/2$, such that the eigenvalue of $\hat{S}^2$ is $N(N+2)/4$. There are $N+1$ states labeled by $S_i=-N/2,\dots 0 \dots N/2$. 

\begin{figure}[t!]
    \centering    \includegraphics[width=1\linewidth]{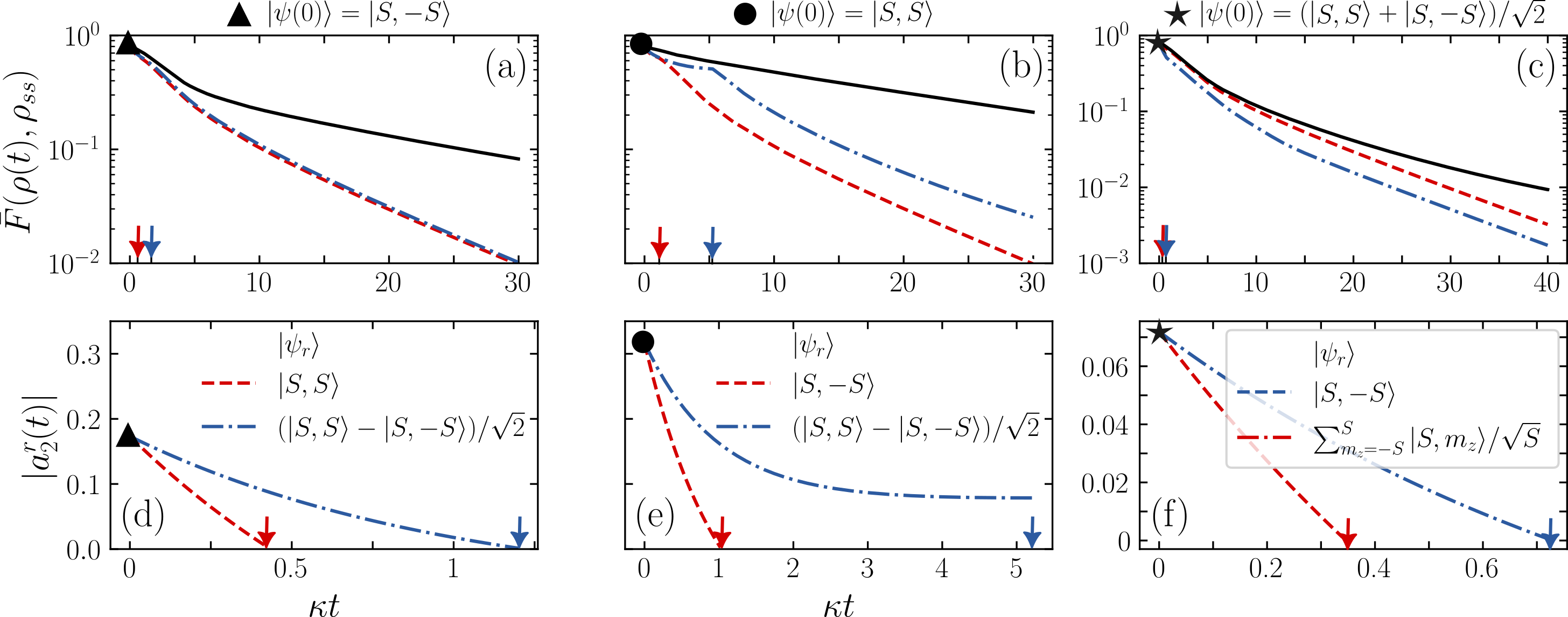}
    \caption{Panels (a) ,(b) and (c) report the distance measure $\bar{F}(\rho(t),\rho_{\text{ss}})$ in Eqs.~\eqref{eq:bar_fidelity} and \eqref{eq:fidelity} as a function of the rescaled time $\kappa t$. Different initial states $\rho=\ket{\psi_0}\bra{\psi_0}$ are considered and reported above the panels. These states $\ket{S,S_i}$ are labeled with the quantum number $S$  of the total spin $\hat{S}^2$ and that $S_i$ of its $z$ component $\hat{S}_z$. We take $S=N/2$ for total spin quantum number. Namely, $S_i \in [-S,S]$, with $S=N/2$. Reset states are reported in the legend, with the associated curves in color-dashed style. Reset-free dynamics is shown in black-solid for comparison. In panels (d), (e), and (f), we plot the modified overlap $a_2^r(t)$ in Eq.~\eqref{eq:akr} as a function of the rescaled time $\kappa t$ for the same resets states considered in panels (a), (b) and (c), respectively. Parameters are set as: $\omega_c/\omega=\kappa/\omega=g/\omega=1$ and $N=20$, while arrows point at time $t_r$ after which the system relaxes according to the reset-free dynamics.}
\label{fig:dicke_reset}
\end{figure}

We now consider the effect of resetting on the dynamics of the Dicke model. We first consider the initial state $\ket{\psi_0}=\ket{S,-S}$, and the effect of the following two different reset states: the maximum spin eigenstate $\ket{\psi_r^0}=\ket{S,S}$ and a coherent superposition of extremal spin eigenstates $\ket{\psi_r^1}=(\ket{S,S}-\ket{S,-S})/\sqrt{2}$. The resulting dynamics is shown in Fig.~\ref{fig:dicke_reset}(a) and (d). A strong Mpemba effect (SME) is observed for both reset states. The coefficient $a_2^r(t_r)$ vanishes at the corresponding reset time $t_r$, as shown in Fig.~\ref{fig:dicke_reset}(d). Consequently, an exponential speedup towards the steady state is evident in  Fig.~\ref{fig:dicke_reset}(a), where the slope of the distance measure $\bar{F}(\rho(t),\rho_{\text{ss}})$ changes at time $t_r$ following the reset protocol.

Next, we consider a different initial state $\ket{\psi_0}=\ket{S,S}$, along with two reset states: the lowest spin eigenstate $\ket{\psi_r^0}=\ket{S,-S}$ and the coherent superposition of extremal spin eigenstates $\ket{\psi_r^1}=(\ket{S,S}-\ket{S,-S})/\sqrt{2}$. The associated results are reported in Figs.~\ref{fig:dicke_reset}(b) and (e). In this case, a SME is observed only for the $\ket{\psi_r^0}$ reset states, whereas a WME is found for $\ket{\psi_r^1}$.

In Figs.~\ref{fig:dicke_reset}(c) and (f), we eventually consider a coherent initial state $\ket{\psi_0}=(\ket{S,S}+\ket{S,-S})/\sqrt{2}$ in the spin eigenbasis. We take the following reset states: the lowest spin eigenstate $\ket{\psi_r^0}=\ket{S,-S}$ and a state $\ket{\psi_r^1}= \sum_{m_z=-S}^S\ket{S,m_z}\sqrt{S}$ equal superposition of all the basis eigenstates $\ket{S,m_z}$. 
We find SME for both the reset states. 
Interestingly, quantum coherence in the reset state $\ket{\psi_r^2}$ accelerates relaxation compared to the incoherent state $\ket{\psi_r^1}$. This result is surprising since the stationary state is mixed and incoherent for the atom-only description \eqref{eq:dicke} of the Dicke model, and it shows, similar to the qubit case, that coherent reset states can be beneficial in accelerating relaxation processes.

\begin{figure}[t!]
    \centering    \includegraphics[width=0.8\linewidth]{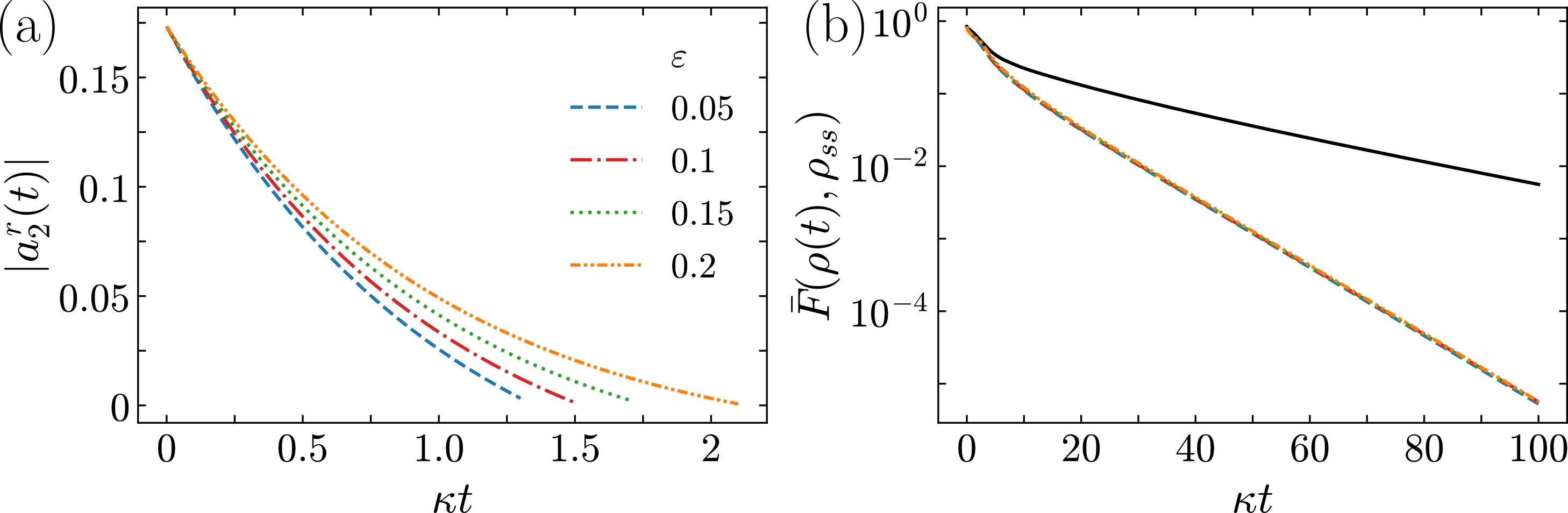}
    \caption{\textcolor{black}{Panel (a) displays the modified overlap $a_2^r(t)$ defined in Eq.~\eqref{eq:akr} as a function of the rescaled time $\kappa t$ for different reset states $\ket{\psi_r^1}=(\sqrt{1-\varepsilon}\ket{S,S}-\sqrt{1+\varepsilon}\ket{S,-S})/\sqrt{2}$, parameterized by $\varepsilon$. The values of $\varepsilon$ are reported in the legend. In all the cases $a_2^r(t)$ can be exactly nullified for $\varepsilon \lesssim 0.2$. The value of $\varepsilon$ only impacts on the time $t_r$ when $a_2^r(t_r)=0$. Panel (b) illustrates that, for the same values of $\varepsilon$ as in panel (a), a strong Mpemba effect persists. Solid line represents the reset-free case (analogous as in Fig.~\ref{fig:dicke_reset}), while dashed colored lines the dynamics with transient resetting up to time $t_r$. The plot indicates that fine-tuned implementation of the reset state is not necessary in our protocol in order to induce relaxation speedup. The parameters are chosen as $\omega_c/\omega=\kappa/\omega=g/\omega=1$ and $N=20$.} 
}
\label{fig:dicke_reset_error}
\end{figure}

\textcolor{black}{We eventually here show that realizing the Mpemba effect via resetting does not rely on fine-tuning the reset state. This is especially relevant in light of implementation of the present protocol in state-of-the-art quantum hardwares, where noise likely lead to an imperfect preparation of the reset state. We therefore consider the initial state $\ket{\psi_0}=\ket{S,-S}$, as in Figs.~\ref{fig:dicke_reset}(a,d), and consider resetting to a deformed state 
\begin{equation}
\ket{\psi_r^1}(\varepsilon)=(\sqrt{1-\varepsilon}\ket{S,S}-\sqrt{1+\varepsilon}\ket{S,-S})/\sqrt{2}, 
\end{equation}
where $\varepsilon$ quantifies the deviation from the ideal reset state $\ket{\psi_r^1}(\varepsilon=0)$. 
As shown in Fig.~\ref{fig:dicke_reset_error}, for small, but finite, values of $\varepsilon \lesssim 0.2$, a strong Mpemba effect persists, similar to that observed in Figs.~\ref{fig:dicke_reset}(a,d) for the $\varepsilon=0$ case. Imperfect preparation of the reset state thus only change the stopping time $t_r$.
This indicates that the emergence of strong Mpemba effect is a robust feature of the protocol with respect to noisy implementations of the reset state.}

\end{document}